\documentclass[11pt, a4paper]{article}
\pdfoutput=1
\usepackage{jheppub}

\usepackage[T1]{fontenc}
\usepackage[babel=true]{microtype}
\usepackage[english]{babel}
\usepackage[dvipsnames]{xcolor}   

\usepackage{mathtools, mathrsfs, dsfont, xfrac, bm}
\usepackage{physics, siunitx, gensymb}
\AtBeginDocument{\RenewCommandCopy\qty\SI}  

\usepackage{array, multirow, multicol, comment}
\usepackage{float}
\usepackage[labelfont=bf, font=small]{caption}
\usepackage{subcaption}
\usepackage{booktabs}

\usepackage{url}
\newcommand{\ms}{\scriptscriptstyle}
\newcommand{\Ucal}{\mathcal{U}}
\newcommand{\Utilde}{\tilde{\mathcal{U}}}
\newcommand{\Uhat}{\hat{\mathcal{U}}}
\newcommand{\DeltaM}[1]{\omega_{#1}}
\newcommand{\msin}[1]{\ensuremath{\mathrm{s}_{#1}}}
\newcommand{\mcos}[1]{\ensuremath{\mathrm{c}_{#1}}}
\newcommand{\mtwosin}[1]{\ensuremath{\mathrm{s}_{2\theta_{#1}}}}
\newcommand{\mtwocos}[1]{\ensuremath{\mathrm{c}_{2\theta_{#1}}}}
\newcommand{\prob}[2]{\ensuremath{P_{#1 #2}}}

\newcommand{\rotmat}[1]{\ensuremath{\mathcal{O}_{#1}}}

\title{\boldmath Exact and hybrid neutrino oscillation probabilities via the evolution operator}

\author[a]{Juan Carlos D'Olivo,}
\author[a,1]{Jos\'{e} Arnulfo Herrera Lara,\note{Corresponding author.}}
\author[b]{and Oscar A. Sampayo}

\affiliation[a]{Instituto de Ciencias Nucleares,
  Universidad Nacional Aut\'{o}noma de M\'{e}xico,\\
  Apartado Postal 70-543, 04510 Ciudad de M\'{e}xico, M\'{e}xico}
 
\affiliation[b]{Departamento de F\'{i}sica,
  Universidad Nacional de Mar del Plata,\\
  Funes 3350, (7600) Mar del Plata, Argentina}

\emailAdd{dolivo@nucleares.unam.mx}
\emailAdd{fis\_pp@ciencias.unam.mx}
\emailAdd{sampayo@mdp.edu.ar}

\abstract{
We revisit the three-flavor neutrino oscillations in constant-density matter using the evolution matrix expressed in terms of the Lagrange-Sylvester interpolation. Taking advantage of Vieta's formulas for the Hamiltonian of the system, we  derive exact closed expressions for the oscillation probabilities $P_{ee}$, $P_{\mu e}$, and $P_{\mu\tau}$, which are concise and simple, and structurally identical to their vacuum counterparts. The CP-violation contribution, proportional to the Jarlskog invariant $\mathcal{J}$, appears as a single, clearly separated term. Our approach does not require introducing neutrino mixing in the medium and avoids using a matter-dependent CP-violation phase, facilitating the separation of genuine CP violation from that induced by matter effects. We explicitly verify that, in the limit $V \to 0$, our expressions for the flavor transition probabilities reduce to the well-known vacuum formulas. The exact structure of the probabilities is preserved if the exact, not very enlightening, trigonometric eigenvalues are replaced by approximate expressions in radicals.
We illustrate this hybrid approach with eigenvalues obtained from two successive rotations of the Hamiltonian. For an experiment like DUNE, the hybrid probability values match those of the exact probabilities within absolute errors of about $2\times 10^{-6}$ in all channels and relative errors less than $0.01 \%$ for $P_{ee}$, $P_{\mu e}$, and $P_{\mu\tau}$.}


\begin{document} 
\maketitle
\flushbottom

\section{Introduction}
\label{intro}
Neutrino oscillations are a well-established, extensively studied phenomenon that conclusively demonstrates that neutrinos are mixed, massive particles \cite{Blennow2013, ParticleDataGroup:2026mpi}. The current experimental and observational data can be interpreted in terms of the minimal extension of the Standard Model, where the three known flavor eigenstates $\ket{\nu_{\alpha}} \ (\alpha = e, \mu, \tau)$ that participate in weak interactions are linear combinations of an equal number of mass eigenstates $\ket{\nu_{i}} \ (i=1,2,3)$, via the relation $\ket{\nu_{\alpha}} = \sum_{i} U^{\ast}_{\alpha i} \ket{\nu_{i}}$. The coefficients $U_{\alpha i}$ are the elements of the Pontecorvo-Maki-Nakagawa-Sakata (PMNS) mixing matrix $U = \rotmat{23} \Gamma \rotmat{13} \Gamma^{\ast} \rotmat{12}$, where $\mathcal{O}_{i j}$ are orthogonal matrices representing rotations by angles $\theta_{ij}\!\in\![0, \pi/2]$ in the respective planes, and $\Gamma = \text{diag}(1, 1, e^{i\delta})$, with $\delta\!\in\![0, 2\pi)$, accounts for CP violation in the leptonic sector. Additionally, the oscillations between the active flavor neutrinos are described by two squared-mass differences: $\Delta m^{2}_{21} \equiv m^{2}_{2} - m^{2}_{1}$ and $\Delta m^{2}_{31} \equiv m^{2}_{3} - m^{2}_{1}$.
 
Most oscillation parameters are known to better than 5\% precision thanks to global analyses of data from solar, atmospheric, reactor, and accelerator neutrino experiments carried out during the last two decades \cite{Esteban:2024eli}. Three main unknowns remain: the sign of $\Delta m^{2}_{31}$, the octant of $\theta_{23}$, and the value of $\delta$. The current global fit is consistent with CP conservation within $1\sigma$ for normal ordering, whereas for inverted ordering values of $\delta$ around $3\pi/2$ are favored over CP conservation at more than $3.6\sigma$ \cite{Esteban:2024eli}. T2K was the first experiment to exclude values of $\delta$ at $3\sigma$, ruling out a large range around $\delta = +\pi/2$ \cite{Abe2020}. A conclusive answer is expected from a new generation of experiments, some already taking data and others under construction, that are designed to address the aforementioned questions and will also contribute to other aspects of neutrino physics and astrophysics \cite{DUNE:2020jqi, DUNECollaboration2025, HyperKamiokande:2018ofw, JUNO:2015zny}. To this end, they will perform high-precision measurements of neutrino oscillations from different sources; JUNO, for instance, began operation in 2025 and has already delivered the most precise determination of the solar oscillation parameters \cite{Abusleme2026}. Proper analysis of these measurements will, in turn, require increasingly precise determinations of event rates. While the calculations can be performed numerically with the required accuracy, having better analytical expressions remains useful for extracting as much information as possible and gaining a deeper understanding of the underlying physics.
 
When neutrinos propagate through a medium, coherent forward scattering with background electrons differs for $\nu_{e}$ and $\nu_{\mu,\tau}$, introducing a density-dependent potential energy in the Hamiltonian that governs the evolution of flavor amplitudes along the neutrino path \cite{Wolfenstein:1977ue, Mikheyev:1985zog, Barger1980}. Consequently, flavor transformations in matter differ non-trivially from those in vacuum, and the expressions of the corresponding probabilities are more difficult to derive \cite{Barger1980, Kuo1989}. The problem of neutrino oscillations in matter cannot be solved exactly for an arbitrary position-dependent density. In the simplest case of two-neutrino mixing, useful expressions for the crossing probability have been derived from exact solutions for linear and exponential density profiles \cite{Toshev:1987jw, Petcov:1987zj}. For these profiles, exact solutions have also been obtained for three neutrinos, but they are given in terms of generalized hypergeometric functions \cite{TorrenteLujan1996, Lehmann2001, Osland2001} and are not easy to handle or interpret physically.
 
An important case that admits exact analytical solutions is the problem of neutrino oscillations in a medium that, to a good approximation, can be considered uniform. The complete solution for three flavors was presented in 1988 by Zaglauer and Schwarzer \cite{Zaglauer1988} in terms of the energy eigenvalues and mixing parameters (angles and CP phase) in matter. Subsequently, a formulation independent of the specific mixing parametrization was given by also diagonalizing the Hamiltonian in matter \cite{XING2000327}. One drawback of such solutions is that the mixing quantities in matter have a rather complicated dependence on the vacuum parameters, making it difficult, in particular, to extract information on CP-violation effects. In 2002, Kimura, Takamura, and Yokomakura \cite{Kimura2002a, Kimura2002} presented simpler exact analytical solutions for all oscillation channels, directly considering the products of the elements of the mixing matrix in matter that appear in the flavor transition amplitudes. More recently, Denton, Parke, and Zhang \cite{Denton2020} gave very neat formulas for these products in terms of the eigenvalues of the Hamiltonian and its principal minors, with the benefit that the oscillation probabilities in matter follow from those in vacuum.
 
An alternative approach that does not involve the mixing parameters in matter is to avoid the diagonalization of the Hamiltonian by directly using the evolution operator in the flavor basis. In a uniform medium, this operator is given by the exponential of a constant $3 \times 3$ matrix, which, thanks to the Cayley-Hamilton theorem, can be expressed as a second-order polynomial in that matrix. There exist different methods to determine the coefficients of the polynomial, such as the Lagrange interpolation formula referred to in the pioneering 1980 work by Barger \emph{et al.}~\cite{Barger1980} and the method implemented by Ohlsson and Snellman~\cite{Ohlsson2000a, Ohlsson2000} for real $U$.
 
Due to the complexity of exact solutions, many studies over the years have addressed the problem by means of perturbative calculations, employing series expansions in the matter potential or in small parameters, such as $\msin{13} = \sin\theta_{13}$ and/or the mass-hierarchy parameter $\Delta m^{2}_{21}/\Delta m^{2}_{31}$ \cite{Kuo1987, Cervera2000, Freund2001, Akhmedov2004, Asano2011, Agarwalla2014, Li2016, Minakata2016, Denton2016}. In Ref.~\cite{Barenboim:2019pfp}, the authors examined how well the various expansions work by comparing them with each other and with the exact solution. Unitarity is not usually satisfied automatically by oscillation formulas derived from standard perturbation theory; rather, it must be imposed order by order in the expansion parameter. While approximate expressions are useful for a better understanding of the problem, they are not without limitations: they have a restricted range of applicability, in the limit of zero potential they do not always reproduce the complete formulas for neutrino oscillations in vacuum, and sometimes they lack the desired simplicity.
 
In this work, we re-examine the problem of three-neutrino oscillations in matter with constant density in terms of the evolution operator in the flavor basis, thus avoiding the introduction of the complicated mixing angles in matter. As in Ref.~\cite{Barger1980}, our starting point is the Lagrange-Sylvester formula for the exponential evolution matrix, but instead of approximate solutions, we seek the exact analytical solution. As we will show, Vieta's formulas, expressed in terms of the traces of the Hamiltonian, are a very convenient way to simplify the calculations. Since only two of the oscillation probabilities, say $\prob{\mu}{e}$ and $\prob{\mu}{\tau}$, are independent \cite{Akhmedov2004}, we derive explicit formulas for both and use them to determine the remaining probabilities via the unitarity condition. Our expressions possess an elegant, systematic, vacuum-like structure, which facilitates their application to both exact and approximate analyses. At $V = 0$, they reduce to the formulas for flavor transitions in vacuum. Furthermore, as a way of corroborating their correctness, we have also verified (not shown) that when expanding them to first order in $\msin{13}$ or in the mass hierarchy, the results coincide exactly with those of Ref.~\cite{Akhmedov2004}, which were derived via series expansions of the evolution operator.
 
As is well known, for three neutrinos propagating in a medium, the exact energy eigenvalues are the real roots of a cubic equation, given by unenlightening trigonometric formulas. Our probabilities, however, depend on the matter effects only through these eigenvalues and the entries of the Hamiltonian, so their structure is preserved when the eigenvalues are replaced by approximate ones. This leads to a hybrid approach: exact formulas for the probabilities evaluated with approximate eigenvalues. We illustrate it with approximate eigenvalues obtained from two successive rotations of the Hamiltonian, following Ref.~\cite{Denton2016}, and show that, for a DUNE-like configuration, the resulting expressions reproduce the exact probabilities with absolute errors of at most about $2\times 10^{-6}$ in all channels, for both neutrinos and antineutrinos, corresponding to relative errors below $10^{-2}\%$ in $P_{ee}$, $P_{\mu e}$, and $P_{\mu\tau}$. In these channels, the absolute errors are at least two orders of magnitude smaller than those of the zeroth-order two-rotation probabilities, which use the same eigenvalues. Furthermore, since these approximate eigenvalues reduce to the vacuum energies of the mass eigenstates for $V = 0$, our flavor transition probabilities reproduce the complete vacuum expressions in that limit, even when written in terms of the approximate eigenvalues.
 
The article is organized as follows. In Section~\ref{Section2}, we introduce the evolution-operator formalism for a potential symmetric about the midpoint of the neutrino path and, as a first application, derive $\prob{\mu}{e}$ and $\prob{\mu}{\tau}$ in vacuum, which also allows us to introduce our notation. Having the vacuum expressions at hand, we can check the matter results by setting the charged-current potential to zero; although straightforward, this check is rarely shown explicitly in the literature. In Section~\ref{matter_sec}, we solve the problem of neutrino oscillations in matter of constant density, explicitly deriving each of the six independent components of the evolution operator in simplified form. These results are used in Section~\ref{ProbaSection} to obtain exact expressions for $\prob{e}{e}$, $\prob{\mu}{e}$, $\prob{\mu}{\tau}$, and the remaining channels. In Section~\ref{hybridsec}, we illustrate the hybrid approach with approximate eigenvalues and assess its performance for a DUNE-like configuration. Section~\ref{Conclusions} contains a discussion of our results and the conclusions. Technical details are collected in Appendix~\ref{AppA}, the explicit verification of the $V \to 0$ limit is given in Appendix~\ref{AppB}, and the derivation of the approximate eigenvalues is presented in Appendix~\ref{AppC}.

\section{Flavor evolution in a symmetric potential}
\label{Section2}
Let us consider neutrinos of a given flavor produced at time $t_0$. In terms of the evolution operator $\Uhat(t, t_0)$ the state of the system $\ket{\psi(t)}$ at time $t \geq t_0$ can be expressed as $\ket{\psi(t)}= \Uhat(t, t_0) \ket{\psi(t_{0})}$, where $\ket{\psi(t_{0})} = \ket{\nu_{\alpha}}$, with $\alpha = e, \mu , \tau$. Expanding $\ket{\psi(t)}$ in terms of the flavor basis, we have\footnote{We have added a superscript to emphasize that these are the amplitudes for an initial state with a given flavor $\alpha$.}
\begin{equation}
\ket{\psi(t)} = \sum\limits_{\beta = e, \mu, \tau} \psi^{(\alpha)}_{\beta}(t) \ket{\nu_{\beta}},
\end{equation}
with $\psi^{(\alpha)}_{\beta}(t) = \matrixel{\beta}{\Uhat(t, t_0)}{\alpha} = \Ucal_{\beta\alpha}(t, t_0)$ and $\psi^{(\alpha)}_{\beta}(t_{0}) = \delta_{\beta\alpha}$.

For typical energies, neutrinos are ultra-relativistic particles and, following the standard approach, we introduce the position $r = t $ $(\hbar=c=1)$ along the neutrino path. Then, the probability $\prob{\alpha}{\beta}(r) = |\psi^{(\alpha)}_{\beta}(r)|^{2}$ to have a state of flavor $\beta$ at a distance $L = r - r_{0}$ from the production point $r_{0}$ can be expressed as
\begin{equation}
\label{oscprob1}
\prob{\alpha}{\beta}(r) = \abs{\mathcal{U}_{\beta\alpha}(r, r_{0})}^{2}.
\end{equation}
The quantities $\Ucal_{\beta\alpha}(r, r_{0})$ are elements of the $3 \times 3$ matrix $\Ucal(r, r_{0})$ representing the evolution operator in the flavor basis:
\begin{equation}
\label{Ev_Op1}
\mathcal{U}(r, r_{0}) = 
 \begin{pmatrix} 
  \Ucal_{e e} & \Ucal_{e \mu} & \Ucal_{e \tau} \\
  \Ucal_{\mu e} & \Ucal_{\mu \mu} & \Ucal_{\mu \tau} \\
  \Ucal_{\tau e} & \Ucal_{\tau \mu} & \Ucal_{\tau \tau} \\
 \end{pmatrix}, 
\end{equation}
where for notational brevity we have omitted the arguments of the matrix elements.
This matrix obeys the differential equation
\begin{equation}
\label{eq:Scheq}
i \pdv{r} \, \Ucal(r, r_{0}) = H(r) \, \Ucal(r, r_{0}),\, \hspace{0.5cm} \Ucal(r_0,r_0) = I, 
\end{equation}
where $H(r)$ is the effective Hamiltonian that governs the evolution of the flavor amplitudes. For neutrinos of energy $E$ propagating in normal matter ($e, p, n$), after subtracting a global phase, it takes the form
\begin{equation}
\label{eq:Hmatter}
H(r) = U H_{0} U^{\dagger} + V(r)Y, 
\end{equation}
where $H_{0} = \text{diag}(0, \Delta_{21}, \Delta_{31})$, $Y = \text{diag}(1, 0, 0)$, and $V(r) = \sqrt{2} G_{\ms\mathrm{F}} n_{e}(r)$. The first term in the previous equation determines the flavor evolution in the vacuum, while the potential $V(r)$ represents the net effect of the coherent interactions of neutrinos with the background particles. Here, $G_{\ms\mathrm{F}}$ is the Fermi constant and $n_{e}(r)$ is the number density of electrons along the neutrino path. For the antineutrinos one has to replace $V(r)$ with $-V(r)$ and $U$ with $U^{\ast}$ (i.e., $\delta$ with $-\delta$) \cite{Meloni2009}.

Since the matrices $\rotmat{23}$ and $\Gamma$ commute with $Y$, we can write
\begin{equation}
\label{redevelop}
\Ucal(r, r_{0}) = \rotmat{23} \Gamma \, \Utilde(r, r_{0}) \Gamma^{\ast} \rotmat{23}^{\mathtt{t}},
\end{equation}
where $\rotmat{23}^{\mathtt{t}}$ is the transpose of $\rotmat{23}$ and
\begin{equation}
\label{Ev_Op2}
\Utilde (r, r_0)=
\begin{pmatrix}
 \Utilde_{ee} & \Utilde_{ea} & \Utilde_{eb} \\
 \Utilde_{ae} & \Utilde_{aa} & \Utilde_{ab} \\
 \Utilde_{be} & \Utilde_{ba} & \Utilde_{bb} \\ 
\end{pmatrix}
\end{equation}
is the matrix representing the evolution operator in an intermediate basis between those of the flavor and mass eigenstates. To find $\Utilde (r, r_0)$, we have to solve a Schrödinger equation like the one in Eq.~\eqref{eq:Scheq} with $H(r)$ replaced by
\begin{equation}
\label{Ham_matter2}
\tilde{H}(r) = \Gamma^{\ast} \rotmat{23}^{\mathtt{t}} H(r) \rotmat{23} \Gamma = \mathcal{O} H_{0} \mathcal{O}^{\mathtt{t}} + V(r)Y,
\end{equation}
with $\mathcal{O} = \rotmat{13} \rotmat{12}$.

Although $\tilde{H}(r)$ is real and symmetric, $\Utilde$ will not be a symmetric matrix unless the electron density is constant or, more generally, a function symmetric about the middle point $\bar{r}$ of the neutrino trajectory in the medium, i.e., $V(2\bar{r} - r) = V(r)$, for $r_{\mathrm{i}}\leq r \leq r_{\mathrm{f}}$ and $\bar{r} = (r_{\mathrm{f}} + r_{\mathrm{i}})/2$, where $r_{\mathrm{i}}$ and $r_{\mathrm{f}}$ are the initial and final points of the trajectory, respectively. In this case,  $\Utilde(r_{\mathrm{f}}, \bar{r}) = \Utilde^{\mathtt{t}}(\bar{r}, r_{\mathrm{i}})$ and the  evolution operator can be factorized as $\Utilde(r_{\mathrm{f}}, r_{\mathrm{i}}) = \Utilde^{\mathtt{t}}(\bar{r}, r_{\mathrm{i}}) \, \Utilde(\bar{r}, r_{\mathrm{i}})$, which implies that
$
\Utilde^{\mathtt{t}}(r_{\mathrm{f}}, r_{\mathrm{i}}) = \Utilde(r_{\mathrm{f}}, r_{\mathrm{i}}),
$
as required for a symmetric matrix.
Consequently, from \eqref{redevelop}, taking into account that $\Gamma^{\ast}(\delta) = \Gamma(-\delta)$, it is immediate to see that
\begin{equation}
\label{condsimU}
\Ucal^{\mathtt{t}}(r_{\mathrm{f}}, r_{\mathrm{i}}; -\delta) = \Ucal(r_{\mathrm{f}}, r_{\mathrm{i}}; \delta).
\end{equation}

From the matrices in \eqref{Ev_Op1} and \eqref {Ev_Op2}, carrying out the matrix product in \eqref{redevelop}, the entries of ${\mathcal{U}}$ can be written explicitly in terms of those of $\Utilde$. In the case of a symmetric potential $\Utilde_{ae} = \Utilde_{ea}$, $\Utilde_{be} = \Utilde_{eb}$, and $\Utilde_{ba} = \Utilde_{ab}$; thus, we get\footnote{Here and in what follows, $\msin{ij} \equiv \sin\theta_{ij}$, $\mcos{ij} \equiv \cos\theta_{ij}$, $\mtwosin{ij} \equiv \sin 2\theta_{ij}$, $\mtwocos{ij} \equiv \cos 2\theta_{ij}$, and $\mcos{\delta} \equiv \cos\delta$.}
\begin{equation}
\label{Relations2}
\begin{aligned}
\Ucal_{e e} &= \Utilde_{ee},\\
\Ucal_{\mu e} &= \mcos{23} \, \Utilde_{ea} + \msin{23} \, \Utilde_{eb} \, e^{i \delta}, \\
\Ucal_{\tau e} &= -\msin{23} \, \Utilde_{ea} + \mcos{23} \, \Utilde_{eb} \, e^{i \delta}, \\
\Ucal_{e \mu} &= \mcos{23} \, \Utilde_{ea} + \msin{23} \, \Utilde_{eb} \, e^{-i \delta}, \\
\Ucal_{\mu \mu} &= \mathrm{c}^{2}_{2 3} \, \Utilde_{aa} + \msin{23}^{2} \, \Utilde_{bb} + 2\msin{23} \mcos{23} \, \Utilde_{ab} \cos\delta, \\
\Ucal_{\tau \mu} &= \msin{23} \mcos{23} (\Utilde_{bb} - \Utilde_{aa}) + (\mcos{23}^{2} e^{i \delta} - \msin{23}^{2} e^{-i \delta})\Utilde_{ab}, \\
\Ucal_{e \tau} &= -\msin{23} \, \Utilde_{ea} + \mcos{23} \, \Utilde_{eb} \, e^{-i \delta}, \\ 
\Ucal_{\mu \tau} &= \msin{23} \mcos{23} (\Utilde_{bb} - \Utilde_{aa}) + (\mcos{23}^{2} e^{-i \delta} - \msin{23}^{2} e^{i \delta})\Utilde_{ab}, \\
\Ucal_{\tau \tau} &= \msin{23}^{2} \, \Utilde_{aa} + \mcos{23}^{2} \, \Utilde_{bb} - 2 \msin{23} \mcos{23}\,\Utilde_{ab} \cos\delta.
\end{aligned}	
\end{equation}
Examining the previous relations, we see that 
\begin{equation}
\label{T_symm}
\Ucal_{\alpha \beta}(\delta) = \Ucal_{\beta \alpha}(-\delta),
\end{equation}
in accordance with Eq.~\eqref{condsimU}. Then, the probabilities $\prob{\alpha}{\beta}(\delta)$ with $\alpha \neq \beta$ can be obtained from $\prob{\beta}{\alpha}(\delta)$ changing $\delta$ to $-\delta$. In addition, by the conservation of probability 
\begin{equation}
\label{Prob_con}
\sum\limits_{\beta} \prob{\alpha}{\beta} = 1, 
\end{equation}
with $\alpha = e, \mu, \tau$. The above conditions imply that from the total set of nine probabilities, only three of them are independent, say, $\prob{\mu}{e} = \abs{\Ucal_{e \mu}}^{2}$, $\prob{\mu}{\tau} = \abs{\Ucal_{\tau \mu}}^{2}$, and $\prob{\tau}{e} = \abs{\Ucal_{e \tau}}^{2}$; the remaining six can be determined in terms of them using \eqref{T_symm} and \eqref{Prob_con}. However, from the expressions in Eq.~\eqref{Relations2} it is immediate to see that $\Ucal_{\tau e}$ results from $\Ucal_{\mu e}$ by doing $\mcos{23} \to -\msin{23}$ and $\msin{23} \to \mcos{23}$. Consequently, we have corroborated that only two of the probabilities $\prob{\alpha}{\beta}$ are sufficient to solve the entire problem of neutrino oscillations in a medium with a potential symmetric about the midpoint of the neutrino path \cite{Akhmedov2004}. In this work, we choose $\prob{\mu}{e}$ and $\prob{\mu}{\tau}$.

From the formulas given in Eq. \eqref {Relations2} we get
\begin{equation}
\label{Generic_Pmu-e}
\begin{aligned}
\prob{\mu}{e} &= \mcos{23}^{2} \abs*{\Utilde_{ea}}^{\!2} + \msin{23}^{2} \abs*{\Utilde_{eb}}^{\!2} + \mtwosin{23} \! \left[\Re(\Utilde_{ea} \, \Utilde^{\ast}_{eb})\cos\delta - \Im(\Utilde_{ea} \, \Utilde^{\ast}_{eb}) \sin\delta\right],	
\end{aligned}
\end{equation}
and
\begin{equation}
\label{Generic_Pmu-tau}
\begin{aligned}
\prob{\mu}{\tau} &= \dfrac{1}{4} \mtwosin{23}^2 \abs*{\Utilde_{bb} - \Utilde_{aa}}^{\!2} + (1 - \mtwosin{23}^2 \cos^{2}{\delta}) \abs*{\Utilde_{ab}}^{\!2} + \mtwosin{23} \mtwocos{23} \Re[(\Utilde_{bb} - \Utilde_{aa}) \, \Utilde_{ab}^{\ast}] \cos\delta \\
&\phantom{=}+ \mtwosin{23} \! \Im[(\Utilde_{bb} - \Utilde_{aa})\,\Utilde_{ab}^{\ast}] \sin\delta.	
\end{aligned}
\end{equation}
These expressions are of particular interest because of their relationship to intrinsic CP violation effects in the leptonic sector. Thanks to the factorization in Eq.~\eqref{redevelop}, the genuine CP violation linked to the phase $\delta$ is distinguished from that induced by matter effects, which are incorporated through the elements of the matrix $\Utilde$.
From the unitarity condition $\Utilde_{ea} \, \Utilde_{eb}^{\ast} + \Utilde_{aa} \, \Utilde_{ab}^{\ast} + \Utilde_{ab} \, \Utilde_{bb}^{\ast} = 0$ it follows that
\begin{equation}
\label{eq2.16}
\Im[(\Utilde_{bb} - \Utilde_{aa})\,\Utilde_{ab}^{\ast}]  = \Im(\Utilde_{ea} \, \Utilde^{\ast}_{eb}), 
\end{equation}
which implies that the term dependent on $\sin\delta$ cancels out in $\prob{\mu}{\mu} = 1 - \prob{\mu}{e} - \prob{\mu}{\tau}$, as required by the condition in Eq. \eqref {T_symm}. This agrees with the expression for $\prob{\mu}{\mu}$ resulting from 
${\mathcal{U}}_{\mu \mu}$ (see Eq. \eqref{Relations2}):
\begin{equation}
\label{Muon_mu_mu}	
\begin{aligned}
\prob{\mu}{\mu} &= |{\mathcal{U}}_{\mu \mu}|^2 = \mcos{23}^{4} \abs*{\Utilde_{aa}}^{2} + \msin{23}^{4} \abs*{\Utilde_{bb}}^{2} + \tfrac{1}{2} \mtwosin{23}^{2}\!\Re(\Utilde_{aa} \Utilde_{bb}^{\ast}) + \mtwosin{23}^{2} \abs*{\Utilde_{ab}}^{2}\!\cos^{2}{\delta} \\
&\phantom{=}+ 2 \mtwosin{23} \!\left[\mcos{23}^{2} \Re(\Utilde_{aa} \, \Utilde_{ab}^{\ast}) + \msin{23}^{2} \Re(\Utilde_{bb} \, \Utilde_{ab}^{\ast})\right]\!\cos\delta,
\end{aligned}
\end{equation}
which does not contain any odd term in $\delta$.

Let us also note that  the survival probability of the electron neutrino can be written as
\begin{equation}
\label{Generic_Pee}
\prob{e}{e} = 1 - \prob{e}{\mu} -  \prob{e}{\tau} = 1 - \abs*{\Utilde_{ea}}^{\!2} - \abs*{\Utilde_{eb}}^{\!2}, 
\end{equation}
where the probabilities $\prob{e}{\mu}$ and $\prob{e}{\tau}$ can be obtained from $\prob{\mu}{e}$ by the procedure described above. The second equality in \eqref{Generic_Pee} is in agreement with the unitarity of the evolution operator in the intermediate basis and implies that $\prob{e}{e}$ does not depend on $\delta$ nor the mixing angle $\theta_{23}$. Next, we will apply formulas \eqref{Generic_Pmu-e} and \eqref{Generic_Pmu-tau} to derive the flavor oscillation probabilities for neutrinos propagating in a uniform medium.

To conclude this section, as a simple application, we derive the oscillation probabilities in vacuum by means of the formulas \eqref{Generic_Pmu-e} and \eqref{Generic_Pmu-tau}. In this case, the elements of the matrix in \eqref{Ev_Op2} are:
\begin{equation}
\label{Relations3}
\begin{aligned}
\Utilde_{ee} &= \mcos{13}^{2} \mcos{12}^{2} + \mcos{13}^{2} \msin{12}^{2} e^{-i \phi_{21}} + \msin{13}^{2} e^{-i \phi_{31}}, \\
\Utilde_{ea} &= \mcos{13} \msin{12} \mcos{12} (e^{-i \phi_{21}} - 1), \\
\Utilde_{eb} &= \msin{13} \mcos{13} (-\mcos{12}^{2} - \msin{12}^{2} e^{-i \phi_{21}} + e^{-i \phi_{31}}), \\
\Utilde_{aa} &= \msin{12}^{2} + \mcos{12}^{2} e^{-i \phi_{21}}, \\
\Utilde_{ab} &= \msin{13} \msin{12} \mcos{12} (1 - e^{-i \phi_{21}}), \\
\Utilde_{bb} &= \msin{13}^{2} \mcos{12}^{2} + \msin{13}^{2} \msin{12}^{2} e^{-i \phi_{21}} + \mcos{13}^{2} e^{-i \phi_{31}},
\end{aligned}
\end{equation}
where $\phi_{k j} = \Delta_{k j} L$, $\Delta_{kj} = \Delta m^{2}_{kj}/2E$, and $L$ denotes the neutrino propagation distance. 
The quantities needed to determine $\prob{\mu}{e}$ and $\prob{\mu}{\tau}$ 
can be easily calculated using the formulas \eqref{Identity1}--\eqref{Identity3}. The results are:
\begin{equation}
\begin{aligned}
\label{Uea_sq}
\abs*{\Utilde_{ea}}^{\!2} &= \mcos{13}^{2} \msin{2\theta_{12}}^{2} \sin[2](\dfrac{\phi_{21}}{2}),\\
\abs*{\Utilde_{eb}}^{\!2} &= - \mcos{13}^{2} \msin{13}^{2}\msin{2\theta_{12}}^{2} \sin[2](\dfrac{\phi_{21}}{2}) + \msin{2\theta_{13}}^{2} \mcos{12}^{2} \sin[2](\dfrac{\phi_{31}}{2}) + \msin{2\theta_{13}}^{2} \msin{12}^{2} \sin[2](\dfrac{\phi_{32}}{2}), \\
\abs*{\Utilde_{ab}}^{\!2} &= \msin{13}^{2} \msin{2\theta_{12}}^{2} \sin[2](\dfrac{\phi_{21}}{2}),\\
\abs*{\Utilde_{bb} - \Utilde_{aa}}^{\!2} &= \left[(1 + \msin{13}^{2})^{2} \mcos{2\theta_{12}}^{2} - \mcos{13}^{4}\right] \sin[2](\dfrac{\phi_{21}}{2}) - 2\mcos{13}^{2} \left[(1 + \msin{13}^{2}) \mcos{2\theta_{12}} - \mcos{13}^{2}\right] \sin[2](\dfrac{\phi_{31}}{2}) \\
&\phantom{=}+2\mcos{13}^{2} \left[(1 + \msin{13}^{2}) \mcos{2\theta_{12}} + \mcos{13}^{2}\right] \sin[2](\dfrac{\phi_{32}}{2}), \\
\Re(\Utilde_{ea} \, \Utilde^{\ast}_{eb})& =  \mcos{13}^{2} \msin{13} \msin{2\theta_{12}} \left[\mcos{2\theta_{12}} \sin[2](\dfrac{\phi_{21}}{2}) + \sin[2](\dfrac{\phi_{31}}{2}) - \sin[2](\dfrac{\phi_{32}}{2})\right], \\
\Re[(\Utilde_{bb} - \Utilde_{aa})\Utilde_{ab}^{\ast}] &= \msin{13}\msin{2\theta_{12}} \left[(1 + \msin{13}^{2}) \mcos{2\theta_{12}}\sin[2](\dfrac{\phi_{21}}{2}) - \mcos{13}^{2}\sin[2](\dfrac{\phi_{31}}{2}) + \mcos{13}^{2}\sin[2](\dfrac{\phi_{32}}{2})\right],\\
\Im(\Utilde_{ea} \, \Utilde^{\ast}_{eb}) &= \Im[(\Utilde_{bb} - \Utilde_{aa})\,\Utilde_{ab}^{\ast}]  = \msin{2\theta_{13}} \mcos{13} \,\msin{2\theta_{12}} \sin(\dfrac{\phi_{21}}{2}) \sin(\dfrac{\phi_{31}}{2}) \sin(\dfrac{\phi_{32}}{2}).
\end{aligned}
\end{equation}

Substituting the expressions listed above into Eqs. \eqref{Generic_Pmu-e} and \eqref{Generic_Pmu-tau} we obtain
\begin{equation}
\label{P_mu-e_vacuum}
\begin{aligned}
\prob{\mu}{e} &= \Lambda^{\mu e}_{21} \sin[2](\dfrac{\Delta_{21} L}{2}) + \Lambda^{\mu e}_{31} \sin[2](\dfrac{\Delta_{31} L}{2}) + \Lambda^{\mu e}_{32} \sin[2](\dfrac{\Delta_{32} L}{2}) \\
&\phantom{=}- 8 \mathcal{J} \sin(\dfrac{\Delta_{21} L}{2}) \sin(\dfrac{\Delta_{31} L}{2}) \sin(\dfrac{\Delta_{32} L}{2}),
\end{aligned}
\end{equation}
with
\begin{equation}
\label{Lambda_vac_mue}
\begin{aligned}
\Lambda^{\mu e}_{21} &= \mcos{23}^2 \mcos{13}^2 \msin{2\theta_{12}}^2 - \frac{1}{4} \msin{23}^2 \msin{2\theta_{13}}^2 \msin{2\theta_{12}}^2 + 4 \mcos{2\theta_{12}} \mathcal{J}_{r} \mcos{\delta}, \\
\Lambda^{\mu e}_{31} &= \msin{23}^2 \msin{2\theta_{13}}^2 \mcos{12}^2 + 4 \mathcal{J}_{r} \mcos{\delta}, \\
\Lambda^{\mu e}_{32} &= \msin{23}^2 \msin{2\theta_{13}}^2 \msin{12}^2 - 4 \mathcal{J}_{r} \mcos{\delta}, \\
\end{aligned}
\end{equation}
and
\begin{equation}
\label{P_mu-tau_vacuum}
\begin{aligned}
\prob{\mu}{\tau} &= \Lambda^{\mu \tau}_{21} \sin[2](\dfrac{\Delta_{21} L}{2}) + \Lambda^{\mu \tau}_{31} \sin[2](\dfrac{\Delta_{31} L}{2}) + \Lambda^{\mu \tau}_{32} \sin[2](\dfrac{\Delta_{32} L}{2}) \\
&\phantom{=}+ 8 \mathcal{J} \sin(\dfrac{\Delta_{21} L}{2}) \sin(\dfrac{\Delta_{31} L}{2}) \sin(\dfrac{\Delta_{32} L}{2}), 
\end{aligned}
\end{equation}
with
\begin{equation}
\label{Lambda_vac_mutau}
\begin{aligned}
\Lambda^{\mu \tau}_{21} &= \mtwosin{23}^{2} (\msin{13}^{2} \msin{12}^{2} - \mcos{12}^{2})(\msin{12}^{2} - \msin{13}^{2} \mcos{12}^{2}) + (1 - \mtwosin{23}^{2} \mcos{\delta}^{2}) \msin{13}^{2} \msin{2\theta_{12}}^{2} +4 \mtwocos{23}\!\left({2}{\mcos{13}^{\!-2}} - 1\right)\!\mcos{2\theta_{12}} \mathcal{J}_{r} \mcos{\delta}, \\
\Lambda^{\mu \tau}_{31} &= \mtwosin{23}^2 \mcos{13}^2 (\msin{12}^2 - \msin{13}^2 \mcos{12}^2) - 4\mtwocos{23}\mathcal{J}_{r} \mcos{\delta}, \\
\Lambda^{\mu \tau}_{32} &= \mtwosin{23}^2 \mcos{13}^2 (\mcos{12}^2 - \msin{13}^2 \msin{12}^2) + 4\mtwocos{23}\mathcal{J}_{r} \mcos{\delta}.
\end{aligned}
\end{equation}
We introduced the Jarlskog invariant $\mathcal{J}$ and the reduced Jarlskog invariant $\mathcal{J}_{r}$, defined as \cite{Jarlskog1985, Minakata2016, Denton2022}
\begin{equation}
\begin{aligned}
\label{J_invariant}
\mathcal{J} &= \mathcal{J}_{r} \sin{\delta}, \\
\mathcal{J}_{r} &= \dfrac{1}{8} \sin{2\theta_{23}} \sin{2\theta_{13}} \cos{\theta_{13}} \sin{2\theta_{12}}. 	
\end{aligned}
\end{equation} 
The survival probability of the electron neutrino can be obtained from $\prob{e}{e} = 1 - \prob{e}{\mu} - \prob{e}{\tau}$, with $\prob{e}{\mu}$ and $\prob{e}{\tau}$ derived from \eqref{P_mu-e_vacuum} by the procedure described above, but the most direct way is to substitute the expressions of $\abs*{\Utilde_{ea}}^{\!2}$ and $\abs*{\Utilde_{eb}}^{\!2}$ into the second equality in \eqref{Generic_Pee}. This gives 
\begin{equation}
\prob{e}{e} = 1 + \Lambda^{ee}_{21} \sin[2](\dfrac{\Delta_{21} L}{2}) + \Lambda^{ee}_{31} \sin[2](\dfrac{\Delta_{31} L}{2}) + \Lambda^{ee}_{32} \sin[2](\dfrac{\Delta_{32} L}{2}),
\end{equation}
where
\begin{alignat}{3}
\label{Lambda_Vac_Pee}
\Lambda^{ee}_{21} &= -\mcos{13}^{4} \mtwosin{12}^{2}, &\qquad
\Lambda^{ee}_{31} &= -\mtwosin{13}^{2} \mcos{12}^{2}, &\qquad\text{and }\quad
\Lambda^{ee}_{32} &= -\mtwosin{13}^{2} \msin{12}^{2}.
\end{alignat}

The expressions for all remaining probabilities have the same structure as $\prob{e}{e}$, $\prob{\mu}{e}$, and $\prob{\mu}{\tau}$, which
allows us to write
\begin{equation}
\label{Global2}
\prob{\alpha}{\beta} =\delta_{\alpha \beta} + \sum\limits_{j<k} \Lambda^{\alpha \beta}_{kj} \sin^{2}{\left(\dfrac{\Delta_{kj} L}{2}\right)} 
+ 8 \epsilon_{\alpha\beta\gamma} \mathcal{J} \prod\limits_{j<k} \sin(\dfrac{\Delta_{kj} L}{2}),
\end{equation}
with $\epsilon_{\alpha\beta\gamma} = 1 (-1)$ for any even (odd) permutation of the indices of $\epsilon_{e \mu \tau}$.
After some algebra, it can be shown that
\begin{equation}
\label{Lambda_coeff}
\Lambda^{\alpha \beta}_{kj} = -4 \Re(U^{\ast}_{\alpha k} U_{\beta k} U_{\alpha j} U^{\ast}_{\beta j}),
\end{equation}
where the explicit expressions for the real part of the quartic products of the mixing matrix elements coincide with those in Table 13.1 of Ref. \cite{Giunti:2007ry}. When writing the last term of Eq. \eqref{Global2}, we took into account that the quantities
$ \Im(U^{\ast}_{\alpha k} U_{\beta k} U_{\alpha j} U^{\ast}_{\beta j}) (j < k)$ are all equal, except for the sign. 

Next, we will show that neutrino oscillation probabilities in matter of constant density take the same form as in vacuum, with minor but significant changes.

\section{Three-neutrino oscillations in matter}
\label{matter_sec}
Our goal is to derive exact formulas for the transition probabilities and express them in the simplest possible form. Let us begin by noticing that $\tilde{H} = \mathcal{O} H_{0} \mathcal{O}^{\mathtt{t}} + VY$ can be diagonalized by an orthogonal transformation: $\mathcal{O}^{\mathtt{t}}_{m} \tilde{H} \mathcal{O}_{m} = H_{\mathtt{D}} = \mathrm{diag}(\xi_{1}, \xi_{2}, \xi_{3})$, with $\mathcal{O}^{\mathtt{t}}_{m} \mathcal{O}_{m} = \mathcal{O}_{m} \mathcal{O}^{\mathtt{t}}_{m} = I$, where the constant matrix $\mathcal{O}_{m}$ reduces to $\mathcal{O}$ for $V=0$. 
The matrices $H$ and ${\tilde H}$ are related by a unitary transformation and therefore
have the same eigenvalues. They are the roots of the cubic equation
\begin{equation}
\label{Char_Pol}
\xi^{3} - \mathsf{a} \xi^{2} + \mathsf{b} \xi - \mathsf{c} = 0\,,
\end{equation}
and satisfy Vieta's formulas
\begin{equation}
\label{vietaformulas}
\begin{aligned}
\mathsf{a} & = \xi_{1} + \xi_{2} + \xi_{3} = {\Tr}(\tilde{H}), \\
\mathsf{b} & = \xi_{1} \xi_{2} + \xi_{1} \xi_{3} + \xi_{2} \xi_{3} = \frac{1}{2} \left[\Tr^2(\tilde{H}) - {\Tr}(\tilde{H}^2)\right],\\
\mathsf{c} & = \xi_{1} \xi_{2} \xi_{3} = {\det}(\tilde{H}),
\end{aligned}
\end{equation}
where $\Tr$ and $\det$ denote the trace and determinant of a matrix, respectively. 
All eigenvalues are real and can be represented by trigonometric functions\footnote{In the case of inverted ordering, the eigenvalues satisfy the ordering relation $\xi^{\ms\mathrm{IO}}_{3} < \xi^{\ms\mathrm{IO}}_{1} < \xi^{\ms\mathrm{IO}}_{2}$, which, in vacuum, reduces to $\Delta_{31} < 0 < \Delta_{21}$. To achieve this, the eigenvalues obtained from Eq.~\eqref{eigenvalues} with $\Delta_{31} < 0$ must be reassigned as $\xi^{\ms\mathrm{IO}}_{1} = \xi_{2}$, $\xi^{\ms\mathrm{IO}}_{2} = \xi_{3}$, and $\xi^{\ms\mathrm{IO}}_{3} = \xi_{1}$.}:

\begin{equation}
\label{eigenvalues}
\xi_{k} = \frac{\mathsf{a}}{3} + \frac{2}{3} \sqrt{\mathsf{a}^2 - 3 \mathsf{b}}\cos\!\left[\Phi- \frac{2\pi}{3}(3 - k)\right], \qquad  k = 1, 2, 3
\end{equation}
\begin{equation}
\Phi = \frac{1}{3} \arccos\!\left(\dfrac{2 \mathsf{a}^3 - 9 \mathsf{a} \mathsf{b} + 27 \mathsf{c}}{2(\mathsf{a}^2 - 3 \mathsf{b})^{3/2}}\right). 
\end{equation}
 Since $\tilde{H}$ does not depend on $\theta_{23}$ nor $\delta$, the $\xi_{k}$ also do not depend on these parameters. 
 
Let us now turn our attention to the evolution operator in the intermediate basis. For a constant density, 
\begin{equation} 
\label{Sol1}
\Utilde(L) = {\exp}(-i\tilde{H}\!L),
\end{equation}
where $L = r_{\mathrm{f}} - r_{\mathrm{i}}$ is the distance that neutrinos travel in the uniform medium. Since $\tilde{H}^{\mathtt{t}} = \tilde{H}$, it follows that $\Utilde^{\mathtt{t}}(L) = \Utilde(L)$ and hence $\mathcal{U}^{\mathtt{t}}(L; -\delta) = \mathcal{U}(L; \delta)$, in agreement with Eq. \eqref{condsimU} for the general case of a symmetric potential.
In terms of the Lagrange-Sylvester interpolation formula,  the exponential matrix in the previous equation can be written as:
\begin{equation}
\begin{aligned}
\label{Ev_Op_gen}
{\exp}(-i\tilde{H}\!L)&= \sum_{k=1}^3 e^{-i L \xi_{k}} \prod_{\substack{j=1 \\ j \neq k}}^{3} 
\frac{\tilde{H} -  \xi_j I} {\xi_k - \xi_j}\\
&=\mathsf{F}_{0}  I + \mathsf{F}_{1}\tilde{H}\!L + \mathsf{F}_{2}\tilde{H}^2\!L^2,
\end{aligned}
\end{equation}
where
\begin{equation}
\label{Matrix_sol1}
\begin{aligned}
\mathsf{F}_{0} &= \dfrac{e^{-i L \xi_{1}}}{\Omega}
\left[\xi_{2} \xi_{3} \DeltaM{32} - \xi_{1} \xi_{3} \DeltaM{31} e^{-i \phi_{21}} + \xi_{1} \xi_{2} \DeltaM{21} e^{-i \phi_{31}}\right], \\
\mathsf{F}_{1} &= \dfrac{e^{-i L \xi_{1}}}{\Omega L}	
\left[-(\xi_2 + \xi_3) \DeltaM{32} + (\xi_1 + \xi_3) \DeltaM{31} e^{-i \phi_{21}} - (\xi_1 + \xi_2) \DeltaM{21} e^{-i \phi_{31}}\right],\\
\mathsf{F}_{2} &= \dfrac{e^{-i L \xi_{1}}}{\Omega L^2}	
\left[\DeltaM{32} - \DeltaM{31} e^{-i \phi_{21}} + \DeltaM{21} e^{-i \phi_{31}}\right],
\end{aligned}
\end{equation}
with $\DeltaM{k j} = \xi_{k} - \xi_{j}$, $\phi_{k j} = \DeltaM{k j} L$, and ${\Omega} = \DeltaM{21} \DeltaM{31} \DeltaM{32}$. 
Using the expressions in Eq. \eqref{eigenvalues} and some trigonometric identities, it can be shown that
\begin{equation}
\label{Trifactor}
\Omega = \sqrt{\tfrac{4}{27} \left(\mathsf{a}^{2} - 3\mathsf{b}\right)^{3} - \tfrac{1}{27}\left(2 \mathsf{a}^{3} - 9 \mathsf{a} \mathsf{b} + 27 \mathsf{c}\right)^2}.
\end{equation}
For notational convenience we write
\begin{equation}
\label{Explicit_Htilde}
\tilde{H} = 
\begin{pmatrix}
\mathsf{h}_{ee} & \mathsf{h}_{ea}& \mathsf{h}_{eb}\\
\mathsf{h}_{ea} & \mathsf{h}_{aa} & \mathsf{h}_{ab}\\
\mathsf{h}_{eb} & \mathsf{h}_{ab}& \mathsf{h}_{bb} \\
\end{pmatrix},
\hspace{1cm}
\tilde{H}^2 =
\begin{pmatrix}
\mathsf{h}^{2}_{ee} & \mathsf{h}^{2}_{ea} & \mathsf{h}^{2}_{eb}\\
\mathsf{h}^{2}_{ea} & \mathsf{h}^{2}_{aa} & \mathsf{h}^{2}_{ab}\\
\mathsf{h}^{2}_{eb} & \mathsf{h}^{2}_{ab} & \mathsf{h}^{2}_{bb} \\
\end{pmatrix}, 
\end{equation}
where
\begin{equation}
\begin{aligned}
\mathsf{h}^{2}_{ee} &= (\mathsf{h}_{ee})^{2} + (\mathsf{h}_{ea})^{2} + (\mathsf{h}_{eb})^{2} \\ 
\mathsf{h}^{2}_{aa} &= (\mathsf{h}_{aa})^{2} + (\mathsf{h}_{ea})^{2} + (\mathsf{h}_{ab})^{2} \\ 
\mathsf{h}^{2}_{bb} &= (\mathsf{h}_{bb})^{2} + (\mathsf{h}_{eb})^{2} + (\mathsf{h}_{ab})^{2} \\
\mathsf{h}^{2}_{ea} &= \mathsf{h}_{eb} \mathsf{h}_{ab} + \mathsf{h}_{ea} (\mathsf{h}_{ee}+ \mathsf{h}_{aa}) \\
\mathsf{h}^{2}_{eb} &= \mathsf{h}_{ea} \mathsf{h}_{ab} + \mathsf{h}_{eb}(\mathsf{h}_{ee} + \mathsf{h}_{bb}) \\
\mathsf{h}^{2}_{ab} &= \mathsf{h}_{ea} \mathsf{h}_{eb} + \mathsf{h}_{ab}(\mathsf{h}_{aa}+ \mathsf{h}_{bb}) 
\end{aligned}
\end{equation}
Here,
\begin{gather}
\label{entriesHtilde}		
\mathsf{h}_{ee} = \msin{12}^{2} \Delta_{21} + \msin{13}^{2} \Delta_{ee} + V, \qquad
\mathsf{h}_{aa} = \mcos{12}^{2} \Delta_{21}, \quad
\mathsf{h}_{bb} = \msin{12}^{2} \Delta_{21} + \mcos{13}^{2} \Delta_{ee},\nonumber\\
\mathsf{h}_{ea} = \msin{12} \mcos{12} \mcos{13} \Delta_{21}, \qquad
\mathsf{h}_{eb} = \msin{13} \mcos{13} \Delta_{ee}, \qquad
\mathsf{h}_{ab} = -\msin{12} \mcos{12} \msin{13} \Delta_{21},
\end{gather}
and, as in Ref. \cite{Parke2016}, we have introduced the notation $\Delta_{ee} \equiv \Delta_{31} - \msin{12}^2 \Delta_{21}$. 

Next, we will determine the matrix elements of the evolution operator in Eq. \eqref{Ev_Op_gen}. Let us first consider the off-diagonal elements:
\begin{equation}
\label{U_ea}
\begin{aligned}
\Utilde_{ea} &= \mathsf{F}_{1} L \mathsf{h}_{ea} + \mathsf{F}_{2} L^2\left[\mathsf{h}_{eb} \mathsf{h}_{ab} + \mathsf{h}_{ea}(\mathsf{a} - \mathsf{h}_{bb})\right] \\
&= \frac{e^{-i \xi_{1} L}}{\Omega}\,\mathsf{h}_{ea} \! \left[\DeltaM{32}(\xi_{1} - q_{1}) - \DeltaM{31}(\xi_{2} - q_{1}) e^{-i \phi_{21}} + \DeltaM{21} (\xi_{3} - q_{1}) e^{-i \phi_{31}}\right],\\
\end{aligned}
\end{equation}
\begin{equation}
\label{U_eb}
\begin{aligned}
\Utilde_{eb} & = \mathsf{F}_{1} L \mathsf{h}_{eb} + \mathsf{F}_{2} L^{2} \left[\mathsf{h}_{ea} \mathsf{h}_{ab} + \mathsf{h}_{eb}(\mathsf{a} - \mathsf{h}_{aa})\right] \\
&= \frac{e^{-i \xi_1L}}{\Omega}\,\mathsf{h}_{eb} \! \left[\DeltaM{32} (\xi_1 - q_2) - \DeltaM{31}(\xi_2 - q_2) e^{-i \phi_{21}} + \DeltaM{21} (\xi_3 - q_2) e^{-i \phi_{31}}\right],\\
\end{aligned}
\end{equation}
\begin{equation}
\label{U_ab}
\begin{aligned}
\Utilde_{ab} &= \mathsf{F}_{1} L \mathsf{h}_{ab} + \mathsf{F}_{2} L^2\left[\mathsf{h}_{ea}\mathsf{h}_{eb} + \mathsf{h}_{ab}(\mathsf{a} - \mathsf{h}_{ee})\right] \\
 &= \frac{e^{-i \xi_{1} L}}{\Omega} \, \mathsf{h}_{ab} \! \left[\DeltaM{32} (\xi_1 - q_3)- \DeltaM{31}(\xi_2 - q_3) e^{-i \phi_{21}} + \DeltaM{21}(\xi_3 - q_3) e^{-i \phi_{31}}\right],
\end{aligned}
\end{equation}
where we have used that $\mathsf{h}_{ee} + \mathsf{h}_{aa} + \mathsf{h}_{bb} = \mathsf{a}$ in accord with the
first equality in Eq. \eqref{vietaformulas}, and we have defined
\begin{gather}
\label{q123}
q_{1} = \mathsf{h}_{bb} - \frac{\mathsf{h}_{eb}\mathsf{h}_{ab}}{\mathsf{h}_{ea}} = \Delta_{31}, \qquad
q_{2} = \mathsf{h}_{aa} -\frac{\mathsf{h}_{ea}\mathsf{h}_{ab}}{\mathsf{h}_{eb}} = \mcos{12}^2 \Delta_{21} \dfrac{\Delta_{31}}{\Delta_{ee}}, \nonumber\\
\label{defq123}
q_{3} = \mathsf{h}_{ee} - \frac{\mathsf{h}_{ea}\mathsf{h}_{eb}}{\mathsf{h}_{ab}} = \Delta_{31} + V.
\end{gather}
As seen in Eq. \eqref{Relations2}, to express ${\mathcal{U}}_{\mu \tau}$ in terms of the entries of the matrix $\Utilde$, in addition to $\Utilde_{ab}$, we need the difference between $\Utilde_{bb}$ and $\Utilde_{aa}$. Using $\mathsf{h}_{ee} +\mathsf{h}_{aa} + \mathsf{h}_{bb} = \xi_{1} + \xi_{2} + \xi_{3} = \mathsf{a}$, we get
\begin{equation}
\label{DifUbbUaa}
\begin{aligned}
\Utilde_{bb} - \Utilde_{aa} &= \mathsf{F}_{1} L (\mathsf{h}_{bb} - \mathsf{h}_{aa}) + \mathsf{F}_{2} L^{2}[(\mathsf{h}_{bb})^{2} - (\mathsf{h}_{aa})^{2} + (\mathsf{h}_{eb})^{2} - (\mathsf{h}_{ea})^{2}]\\
&=\frac{e^{-i \xi_{1} L}}{\Omega}\Big\{\!\big[\!-(\xi_{2} + \xi_{3}) \DeltaM{32} + (\xi_{1} + \xi_{3}) \DeltaM{31} e^{-i \phi_{21}} - (\xi_{1} + \xi_{2})\DeltaM{21} e^{-i \phi_{31}}\big] \\
&+\big(\DeltaM{32} - \DeltaM{31} e^{-i \phi_{21}} + \DeltaM{21} e^{-i \phi_{31}}\big)\big(\mathsf{h}_{bb} + \mathsf{h}_{aa} + \mathsf{h}_{ee} - q_{4}\big)\Big\} (\mathsf{h}_{bb} - \mathsf{h}_{aa}) \\
&= \frac{e^{-i \xi_{1} L}}{\Omega}(\mathsf{h}_{bb} - \mathsf{h}_{aa})\!\big[\DeltaM{32}(\xi_{1} - q_{4}) - \DeltaM{31}(\xi_{2} - q_{4}) e^{-i \phi_{21}} + \DeltaM{21}(\xi_{3} - q_{4})e^{-i \phi_{31}}\big],	
\end{aligned}
\end{equation}
where
\begin{equation}
\label{defq4}
q_{4} = \mathsf{h}_{ee} - \frac{(\mathsf{h}_{eb})^{2} - (\mathsf{h}_{ea})^2}{\mathsf{h}_{bb} - \mathsf{h}_{aa}} = \dfrac{\Delta_{21} \Delta_{31}}{\Delta_{\mu\tau}}(\msin{12}^{2} - \mcos{12}^{2} \msin{13}^{2}) + V,	
\end{equation}
and, for convenience, we have defined
\begin{equation}
\label{defD}
\Delta_{\mu\tau} \equiv \mcos{13}^{2} \Delta_{ee} - \mtwocos{12} \Delta_{21} = \mcos{13}^{2} \Delta_{31} -(\mcos{12}^{2} - \msin{12}^{2} \msin{13}^{2}) \Delta_{21}.
\end{equation}
Therefore, $\Utilde_{bb} - \Utilde_{aa}$ acquires an expression similar to those of the non-diagonal entries. 
This is advantageous for the analytical calculation of flavor transition probabilities and is the reason why 
we selected $\prob{\mu}{\tau}$ as the probability to be computed.  

In this way, we have gathered all the necessary elements to calculate the transition probabilities between the three neutrino flavors. The survival probabilities can be obtained from them and the probability conservation; even so, it is convenient to supplement this section by providing formulas for the flavor survival amplitudes, in which case the term proportional to the identity in Eq. \eqref{Ev_Op_gen} also contributes. Specifically, the steps to calculate $\Utilde_{ee}$ are as follows:
\begin{equation}
\label{U_ee1}
\begin{aligned}
\Utilde_{ee} &= \mathsf{F}_{0} + \mathsf{F}_{1} L \mathsf{h}_{ee} + \mathsf{F}_{2}L^{2} \mathsf{h}^{2}_{ee} \\
&= \frac{e^{-i \xi_{1} L}}{\Omega}\Big\{\DeltaM{32} \left[\xi_{2} \xi_{3} - (\mathsf{a} - \xi_{1})\mathsf{h}_{ee} 
+ (\mathsf{h}_{ee})^{2} + (\mathsf{h}_{ea})^{2} + (\mathsf{h}_{eb})^{2}\right] \\
& \phantom{=} \hspace{1cm}- \DeltaM{31}\left[\xi_{1} \xi_{3} - (\mathsf{a} - \xi_{2})\mathsf{h}_{ee} +  (\mathsf{h}_{ee})^{2} + (\mathsf{h}_{ea})^{2} + (\mathsf{h}_{eb})^{2}\right] e^{-i \phi_{21}}\\
& \phantom{=} \hspace{1cm}+ \DeltaM{21}\left[\xi_{1} \xi_{2} - (\mathsf{a} - \xi_{3})\mathsf{h}_{ee} + (\mathsf{h}_{ee})^{2} + (\mathsf{h}_{ea})^{2} + (\mathsf{h}_{eb})^{2}\right] e^{-i \phi_{31}}\Big\}. \\
\end{aligned}
\end{equation}
Using ${\Tr}(\tilde{H}) = \mathsf{h}_{ee} + \mathsf{h}_{aa} + \mathsf{h}_{bb}$ and
\begin{equation}
\dfrac{1}{2}[\Tr^{2}(\tilde{H}) - {\Tr}(\tilde{H}^2)] = \mathsf{h}_{ee} \mathsf{h}_{aa} + \mathsf{h}_{ee} \mathsf{h}_{bb} + \mathsf{h}_{aa} \mathsf{h}_{bb} - (\mathsf{h}_{ea})^{2} - (\mathsf{h}_{eb})^{2} - (\mathsf{h}_{ab})^{2},
\end{equation}
together with the first two equalities in \eqref{vietaformulas}, we find
\begin{equation}
\label{C1-U_ee2}
\begin{aligned}
\xi_{2} \xi_{3} - (\mathsf{a} - \xi_{1})\mathsf{h}_{ee} + (\mathsf{h}_{ee})^{2} + (\mathsf{h}_{ea})^{2} + (\mathsf{h}_{eb})^{2} 
&= \mathsf{h}_{aa} \mathsf{h}_{bb} - (\mathsf{a} - \mathsf{h}_{ee})\xi_{1} + \xi^{2}_{1} - (\mathsf{h}_{ab})^{2} \\
&= (\mathsf{h}_{aa} - \xi_{1})(\mathsf{h}_{bb} - \xi_{1}) - (\mathsf{h}_{ab})^{2}. \\
\end{aligned}
\end{equation}
Similarly,
\begin{equation}
\label{C2-U_ee2}
\begin{aligned}
\xi_{1} \xi_{3} - (\mathsf{a} - \xi_{2})\mathsf{h}_{ee} + (\mathsf{h}_{ee})^{2} + (\mathsf{h}_{ea})^{2} + (\mathsf{h}_{eb})^{2} 
&= (\mathsf{h}_{aa} - \xi_{2})(\mathsf{h}_{bb} - \xi_{2})  - (\mathsf{h}_{ab})^{2} ,\\ 
\end{aligned}
\end{equation}
and
\begin{equation}
\label{C3-U_ee2}
\begin{aligned}
\xi_{1} \xi_{2} - (\mathsf{a} - \xi_{3})\mathsf{h}_{ee} + (\mathsf{h}_{ee})^{2} + (\mathsf{h}_{ea})^{2} + (\mathsf{h}_{eb})^{2} 
&= (\mathsf{h}_{aa} - \xi_{3})(\mathsf{h}_{bb} - \xi_{3})  - (\mathsf{h}_{ab})^{2}. \\ 
\end{aligned}
\end{equation}
Substituting the equalities in \eqref{C1-U_ee2}--\eqref{C3-U_ee2} into the corresponding square brackets of  Eq. \eqref{U_ee1} yields
\begin{equation}
\label{U_ee2bis}
\begin{aligned}
\Utilde_{ee} &= \frac{e^{-i \xi_{1} L}}{\Omega}\Big\{\DeltaM{32} \left[(\xi_1 - \mathsf{h}_{aa})(\xi_1 - \mathsf{h}_{bb}) - (\mathsf{h}_{ab})^2\right]- \DeltaM{31} \left[(\xi_{2} - \mathsf{h}_{aa})(\xi_{2} - \mathsf{h}_{bb})  - (\mathsf{h}_{ab})^{2}\right] e^{-i \phi_{21}} \\
&\phantom{=}+\DeltaM{21} \left[(\xi_{3} - \mathsf{h}_{aa})(\xi_3 - \mathsf{h}_{bb}) - (\mathsf{h}_{ab})^{2}\right] e^{-i \phi_{31}}\Big\}.
\end{aligned}	 
\end{equation}

Through similar algebraic manipulations, the other diagonal entries of $\Utilde$ become 
\begin{equation}
\begin{aligned}
\Utilde_{aa} &= \frac{e^{-i \xi_{1} L}}{\Omega}\Big\{\DeltaM{32} \left[(\xi_{1} - \mathsf{h}_{ee})(\xi_{1} - \mathsf{h}_{bb}) - (\mathsf{h}_{eb})^{2}\right] - \DeltaM{31} \left[(\xi_{2} - \mathsf{h}_{ee})(\xi_{2} - \mathsf{h}_{bb}) - (\mathsf{h}_{eb})^{2}\right] e^{-i \phi_{21}} \\
&\phantom{=}+\DeltaM{21} \left[(\xi_{3} - \mathsf{h}_{ee})(\xi_{3} - \mathsf{h}_{bb}) - (\mathsf{h}_{eb})^{2}\right] e^{-i \phi_{31}}\Big\},
\end{aligned}
\end{equation}
\begin{equation}
\begin{aligned}
\Utilde_{bb} &= \frac{e^{-i \xi_{1} L}}{\Omega}\Big\{\DeltaM{32} \left[(\xi_{1} - \mathsf{h}_{ee})(\xi_{1} - \mathsf{h}_{aa})  - (\mathsf{h}_{ea})^{2}\right] - \DeltaM{31} \left[(\xi_2 - \mathsf{h}_{ee})(\xi_{2} - \mathsf{h}_{aa}) - (\mathsf{h}_{ea})^{2}\right] e^{-i \phi_{21}} \\
&\phantom{=}+\DeltaM{21} \left[(\xi_{3} - \mathsf{h}_{ee})(\xi_{3} - \mathsf{h}_{aa}) - (\mathsf{h}_{ea})^{2}\right] e^{-i \phi_{31}}\Big\}.
\end{aligned}
\end{equation}
From the last two equations, we see that $\Utilde_{bb} - \Utilde_{aa}$ agrees with the expression in Eq. \eqref{DifUbbUaa}.

Let us also note that
\begin{equation}
\begin{pmatrix}
\mathsf{h}_{\mu\mu} - \xi_{i} & \mathsf{h}_{\mu\tau} \\
\mathsf{h}^\ast_{\mu\tau} & \mathsf{h}_{\tau\tau} - \xi_{i} \\
\end{pmatrix}
=
\begin{pmatrix}
 \mcos{23} & \msin{23} e^{i \delta} \\
-\msin{23} & \mcos{23} e^{i \delta} \\
\end{pmatrix} 
\begin{pmatrix}
\mathsf{h}_{aa} - \xi_{i} & \mathsf{h}_{ab} \\
\mathsf{h}_{ab} & \mathsf{h}_{bb} - \xi_{i} \\
\end{pmatrix}
\begin{pmatrix}
\mcos{23} & -\msin{23} \\
\msin{23} e^{-i \delta} & \mcos{23} e^{-i \delta} \\
\end{pmatrix}, \\
\end{equation} 
with $i = 1,2,3$. From the properties of the determinant, we have
\begin{equation}
\label{detequality}
\begin{vmatrix}
\xi_{i} - \mathsf{h}_{aa} & \mathsf{h}_{ab} \\
\mathsf{h}_{ab} &  \xi_{i} - \mathsf{h}_{bb} \\
\end{vmatrix}
= 
\begin{vmatrix}
\xi_{i} - \mathsf{h}_{\mu\mu} & \mathsf{h}_{\mu\tau} \\
\mathsf{h}^{\ast}_{\mu\tau} &  \xi_{i} - \mathsf{h}_{\tau\tau} \\
\end{vmatrix},
\end{equation}
which allows us to rewrite Eq. \eqref{U_ee2bis} as
\begin{equation}
\label{U_ee3}
\begin{aligned}
\Utilde_{ee} &= \frac{e^{-i \xi_{1} L}}{\Omega}\Big\{\!\DeltaM{32} \! \left[(\xi_{1} - \mathsf{h}_{\mu\mu})(\xi_{1} - \mathsf{h}_{\tau\tau}) - |\mathsf{h}_{\mu\tau}|^{2}\right]	 - \DeltaM{31} \! \left[(\xi_{2} - \mathsf{h}_{\mu\mu})(\xi_{2} - \mathsf{h}_{\tau\tau})  - |\mathsf{h}_{\mu\tau}|^{2}\right] e^{-i \phi_{21}} \\
&\phantom{=}+ \DeltaM{21} \! \left[(\xi_{3} - \mathsf{h}_{\mu\mu})(\xi_{3} - \mathsf{h}_{\tau\tau}) - |\mathsf{h}_{\mu\tau}|^{2}\right] e^{-i \phi_{31}}\!\Big\}.
\end{aligned}
\end{equation}

It should now be noted that the evolution operator in the flavor basis $\mathcal{U}$ can also be determined by the Lagrange-Sylvester interpolation formula and is given by an expression like the one in \eqref{Ev_Op_gen} with $\tilde{H}$ replaced by the Hamiltonian
\begin{equation}
\label{Explicit_H}
H = \begin{pmatrix}
\mathsf{h}_{ee} & \mathsf{h}_{e\mu} & \mathsf{h}_{e\tau}\\
\mathsf{h}^\ast_{e\mu} & \mathsf{h}_{\mu\mu} & \mathsf{h}_{\mu\tau}\\
\mathsf{h}^\ast_{e\tau} & \mathsf{h}^\ast_{\mu\tau} & \mathsf{h}_{\tau\tau} \\
\end{pmatrix}.
\end{equation}
Thus, an expression for $\mathcal{U}_{ee}$ can be obtained following a procedure analogous to the one used to derive the formula for $\Utilde_{ee}$ in Eq. \eqref{U_ee2bis}. The result coincides with the right-hand side of Eq. \eqref{U_ee3}, in agreement with the first equality in Eq. \eqref{Relations2}. Formulas for ${\mathcal{U}}_{\mu\mu}$ and ${\mathcal{U}}_{\tau\tau}$ can be derived in an analogous way. Hence, the survival amplitudes for the three neutrino flavors can be written as follows:
\begin{equation}
\label{DetermForm}
\begin{aligned}
\mathcal{U}_{ee} &= \dfrac{e^{-i \xi_{1} L}}{\Omega} \left(\DeltaM{32} \begin{vmatrix}
\mathsf{h}_{\mu\mu} - \xi_{1} & \mathsf{h}^{\ast}_{\mu\tau} \\
\mathsf{h}_{\mu\tau} & \mathsf{h}_{\tau\tau} - \xi_{1} \\
\end{vmatrix} - \DeltaM{31} \begin{vmatrix}
\mathsf{h}_{\mu\mu} - \xi_{2} & \mathsf{h}^{\ast}_{\mu\tau} \\
\mathsf{h}_{\mu\tau} & \mathsf{h}_{\tau\tau} - \xi_{2} \\
\end{vmatrix} e^{-i \phi_{21}} \right.\\
&\left.\phantom{=} + \DeltaM{21} \begin{vmatrix}
\mathsf{h}_{\mu\mu} - \xi_{3} & \mathsf{h}^{\ast}_{\mu\tau} \\
\mathsf{h}_{\mu\tau} & \mathsf{h}_{\tau\tau} - \xi_{3} \\
\end{vmatrix} e^{-i \phi_{31}}\right),\\
\mathcal{U}_{\mu\mu} &= \dfrac{e^{-i \xi_{1} L}}{\Omega} \left(\DeltaM{32} \begin{vmatrix}
\mathsf{h}_{ee} - \xi_{1} & \mathsf{h}^{\ast}_{e\tau} \\
\mathsf{h}_{e\tau} & \mathsf{h}_{\tau\tau} - \xi_{1} \\
\end{vmatrix} - \DeltaM{31} \begin{vmatrix}
\mathsf{h}_{ee} - \xi_{2} & \mathsf{h}^{\ast}_{e\tau} \\
\mathsf{h}_{e\tau} & \mathsf{h}_{\tau\tau} - \xi_{2} \\
\end{vmatrix} e^{-i \phi_{21}} \right. \\
& \left.\phantom{=}+ \DeltaM{21} \begin{vmatrix}
\mathsf{h}_{ee} - \xi_{3} & \mathsf{h}^{\ast}_{e\tau} \\
\mathsf{h}_{e\tau} & \mathsf{h}_{\tau\tau} - \xi_{3} \\
\end{vmatrix} e^{-i \phi_{31}}\right), \\
\mathcal{U}_{\tau\tau} &= \dfrac{e^{-i \xi_{1} L}}{\Omega} \left(\DeltaM{32} \begin{vmatrix}
\mathsf{h}_{ee} - \xi_{1} & \mathsf{h}^{\ast}_{e\mu} \\
\mathsf{h}_{e\mu} & \mathsf{h}_{\mu\mu} - \xi_{1} \\
\end{vmatrix} - \DeltaM{31} \begin{vmatrix}
\mathsf{h}_{ee} - \xi_{2} & \mathsf{h}^{\ast}_{e\mu} \\
\mathsf{h}_{e\mu} & \mathsf{h}_{\mu\mu} - \xi_{2} \\
\end{vmatrix} e^{-i \phi_{21}} \right. \\
& \left.\phantom{=}+ \DeltaM{21} \begin{vmatrix}
\mathsf{h}_{ee} - \xi_{3} & \mathsf{h}^{\ast}_{e\mu} \\
\mathsf{h}_{e\mu} & \mathsf{h}_{\mu\mu} - \xi_{3} \\
\end{vmatrix} e^{-i \phi_{31}}\right).
\end{aligned}
\end{equation}

In the following section, we will apply the expressions presented previously to derive exact analytical formulas for the flavor transition probabilities between three active neutrinos as a function of the oscillation parameters and the energy of the neutrinos propagating in matter with constant density.
\section{Oscillation probabilities in a uniform medium}
\label{ProbaSection}
Firstly, for the sake of brevity in the expressions presented below, we introduce some notations:
\begin{equation}
\label{defA,B}
\mathscr{A}_{i} = \xi_{i} - \Delta_{31}, \hspace{1cm} \mathscr{B}_{i} = \xi_{i} - \dfrac{\Delta_{21}\Delta_{31}}{\Delta_{ee}}\mcos{12}^{2},
\end{equation}
\begin{equation}
\label{defC,D}
\mathscr{C}_{i} = \xi_{i} - \Delta_{31} - V, \hspace{1cm} \mathscr{D}_{i} = \xi_{i} - \dfrac{\Delta_{21} \Delta_{31}}{\Delta_{\mu\tau}} (\msin{12}^{2} - \mcos{12}^{2} \msin{13}^{2}) - V,
\end{equation}
with $i = 1,2,3$.

To derive an explicit formula for $\prob{\mu}{e}$ we need to calculate $\abs*{\Utilde_{ea}}^{\!2}$, $\abs*{\Utilde_{eb}}^{\!2}$, and the real and imaginary parts of  $\Utilde_{ea} \, \Utilde^{\ast}_{eb}$. Using the equalities given in Appendix~\ref{AppA}, taking into account that $\DeltaM{32} \mathscr{A}_{1}- \DeltaM{31}\mathscr{A}_{2} + \DeltaM{21}\mathscr{A}_{3} = \DeltaM{32} \mathscr{B}_{1} - \DeltaM{31}\mathscr{B}_{2} + \DeltaM{21}\mathscr{B}_{3} = 0$, we find:
\begin{equation}
\begin{aligned}
\label{U_ea1_exact}
\abs*{\Utilde_{ea}}^2 &= 4\frac{(\mathsf{h}_{ea})^2}{\Omega} \left[\dfrac{\mathscr{A}_{2} \mathscr{A}_{1}}{\DeltaM{21}} \sin[2](\frac{\phi_{21}}{2}) - \dfrac{\mathscr{A}_{3} \mathscr{A}_{1}}{\DeltaM{31}} \sin[2](\frac{\phi_{31}}{2}) + \dfrac{\mathscr{A}_{3} \mathscr{A}_{2}}{\DeltaM{32}} \sin[2](\frac{\phi_{32}}{2})\right], \\
\abs*{\Utilde_{eb}}^2 &= 4\frac{(\mathsf{h}_{eb})^2}{\Omega} \left[\dfrac{\mathscr{B}_{2} \mathscr{B}_{1}}{\DeltaM{21}} \sin[2](\frac{\phi_{21}}{2}) - \dfrac{\mathscr{B}_{3} \mathscr{B}_{1}}{\DeltaM{31}} \sin[2](\frac{\phi_{31}}{2}) + \dfrac{\mathscr{B}_{3} \mathscr{B}_{2}}{\DeltaM{32}} \sin[2](\frac{\phi_{32}}{2})\right],
\end{aligned}
\end{equation}
\begin{equation}
\label{Re_P_mu-e_exact}
\begin{aligned}
\Re(\Utilde_{ea} \, \Utilde^{\ast}_{eb}) &= 2 \frac{\mathsf{h}_{ea}\mathsf{h}_{eb}}{\Omega} \left[\dfrac{\mathscr{A}_{2} \mathscr{B}_{1} + \mathscr{A}_{1} \mathscr{B}_{2}}{\DeltaM{21}} \sin[2](\frac{\phi_{21}}{2})\right.\\
& \left.\phantom{=} - \dfrac{\mathscr{A}_{3} \mathscr{B}_{1} + \mathscr{A}_{1} \mathscr{B}_{3}}{\DeltaM{31}}\sin[2](\frac{\phi_{31}}{2})
+ \dfrac{\mathscr{A}_{3} \mathscr{B}_{2} + \mathscr{A}_{2} \mathscr{B}_{3}}{\DeltaM{32}}\sin[2](\frac{\phi_{32}}{2})\right],\\
\Im(\Utilde_{ea} \, \Utilde^{\ast}_{eb}) &= \frac{\mathsf{h}_{ea}\mathsf{h}_{eb}}{\Omega} 
\left[\dfrac{\mathscr{A}_{2}\mathscr{B}_{1} - \mathscr{A}_{1}\mathscr{B}_{2}}{\DeltaM{21}} \sin(\phi_{21})\right.\\
& \left.\phantom{=} - \dfrac{\mathscr{A}_{3} \mathscr{B}_{1} - \mathscr{A}_{1} \mathscr{B}_{3}}{\DeltaM{31}}\sin(\phi_{31})
+ \dfrac{\mathscr{A}_{3} \mathscr{B}_{2} - \mathscr{A}_{2} \mathscr{B}_{3}}{\DeltaM{32}}\sin(\phi_{32})\right]\\
& = \frac{\mathsf{h}_{ea}\mathsf{h}_{eb}}{\Omega}(q_1 - q_2)\Big[\sin(\phi_{21})
-\sin(\phi_{31}) + \sin(\phi_{32})\Big].
\end{aligned}
\end{equation}
Substituting the above expressions in the formula for $\prob{\mu}{e}$ given in Eq. \eqref{Generic_Pmu-e} and considering the trigonometric relation $\sin\phi_{21} - \sin\phi_{31} + \sin\phi_{32} = 4 \sin(\phi_{21}/2) \sin(\phi_{31}/2) \sin(\phi_{32}/2)$, which is valid as long as $\phi_{21} - \phi_{31} + \phi_{32} = 0$, we arrive at
\begin{equation}
\label{P_mu_e}
\begin{aligned}
\prob{\mu}{e} &= \Lambda^{\mu e}_{21}\sin[2](\frac{\DeltaM{21}L}{2}) + \Lambda^{\mu e}_{31} \sin[2](\frac{\DeltaM{31}L}{2}) + \Lambda^{\mu e}_{32} \sin[2](\frac{\DeltaM{32}L}{2})\\ 
&\phantom{=}-8\dfrac{\Delta_{21} \Delta_{31} \Delta_{32}}{\DeltaM{21} \DeltaM{31} \DeltaM{32}} \mathcal{J} \sin(\frac{\DeltaM{21}L}{2}) \sin(\frac{\DeltaM{31}L}{2}) 
\sin(\frac{\DeltaM{32}L}{2}),
\end{aligned}
\end{equation}
where 	
\begin{equation}
\label{LambdaEMU31}
\begin{aligned}
\Lambda^{\mu e}_{21} &= \dfrac{1}{\Omega \DeltaM{21}} \Big[\mcos{23}^2 \mcos{13}^2 \mtwosin{12}^2 \Delta_{21}^2 
\mathscr{A}_{2}\mathscr{A}_{1} + \msin{23}^2 \mtwosin{13}^2 \Delta_{ee}^2\mathscr{B}_{2} \mathscr{B}_{1}
+ 4 \Delta_{21} \Delta_{ee} \big(\mathscr{A}_{2} \mathscr{B}_{1} + \mathscr{B}_{2} \mathscr{A}_{1} \big)\mathcal{J}_{r} \mcos{\delta}\Big],\\
\Lambda^{\mu e}_{31} &= \dfrac{-1}{\Omega \DeltaM{31}} \Big[\mcos{23}^2 \mcos{13}^2 \mtwosin{12}^2 \Delta_{21}^2 
\mathscr{A}_{3}\mathscr{A}_{1} + \msin{23}^2 \mtwosin{13}^2 \Delta_{ee}^2\mathscr{B}_{3}\mathscr{B}_{1}
+ 4 \Delta_{21} \Delta_{ee} \big(\mathscr{A}_{3} \mathscr{B}_{1} + \mathscr{B}_{3} \mathscr{A}_{1}\big)\mathcal{J}_{r} \mcos{\delta}\Big],\\
\Lambda^{\mu e}_{32} &= \dfrac{1}{\Omega \DeltaM{32}} \Big[\mcos{23}^2 \mcos{13}^2 \mtwosin{12}^2 \Delta_{21}^2 
\mathscr{A}_{3} \mathscr{A}_{2} + \msin{23}^2 \mtwosin{13}^2 \Delta_{ee}^2\mathscr{B}_{3} \mathscr{B}_{2}
+ 4 \Delta_{21} \Delta_{ee} \big(\mathscr{A}_{3} \mathscr{B}_{2} + \mathscr{B}_{3} \mathscr{A}_{2} \big)\mathcal{J}_{r} \mcos{\delta}\Big].
\end{aligned}
\end{equation}
Through similar steps, we get,
\begin{equation}
\label{U_ab1_exact}
\begin{aligned}
\abs*{\Utilde_{ab}}^2 &= 4\frac{(\mathsf{h}_{ab})^2}{\Omega} \left[\dfrac{\mathscr{C}_{2} \mathscr{C}_{1}}{\DeltaM{21}} \sin[2](\frac{\phi_{21}}{2}) - \dfrac{\mathscr{C}_{3}\mathscr{C}_{1}}{\DeltaM{31}} \sin[2](\frac{\phi_{31}}{2}) + \dfrac{\mathscr{C}_{3}\mathscr{C}_{2}}{\DeltaM{32}} \sin[2](\frac{\phi_{32}}{2})\right], \\
\abs*{\Utilde_{bb} - \Utilde_{aa}}^2 &= 4\frac{(\mathsf{h}_{bb} - \mathsf{h}_{aa})^2}{\Omega} \left[\dfrac{\mathscr{D}_{2}\mathscr{D}_{1}}{\DeltaM{21}} \sin[2](\frac{\phi_{21}}{2}) - \dfrac{\mathscr{D}_{3} \mathscr{D}_{1}}{\DeltaM{31}} \sin[2](\frac{\phi_{31}}{2}) + \dfrac{\mathscr{D}_{3}\mathscr{D}_{2}}{\DeltaM{32}} \sin[2](\frac{\phi_{32}}{2})\right],
\end{aligned}
\end{equation}
\begin{equation}
\label{DeltaU_exact3}
\begin{aligned}
\Re\big[(\tilde{\mathcal{U}}_{b b} - \tilde{\mathcal{U}}_{a a})\tilde{\mathcal{U}}^*_{a b}\big] &= 2 \frac{(\mathsf{h}_{bb} - \mathsf{h}_{aa})\mathsf{h}_{ab}}{\Omega}\left[\dfrac{\mathscr{D}_{2}\mathscr{C}_{1} + \mathscr{D}_{1} \mathscr{C}_{2}}{\DeltaM{21}} \sin[2](\frac{\phi_{21}}{2})\right.\\
& \left.\phantom{=} -\dfrac{\mathscr{D}_{3} \mathscr{C}_{1} + \mathscr{D}_{1} \mathscr{C}_{3}}{\DeltaM{31}}\sin[2](\frac{\phi_{31}}{2}) + \dfrac{\mathscr{D}_{3} \mathscr{C}_{2} + \mathscr{D}_{2} \mathscr{C}_{3}}{\DeltaM{32}}\sin[2](\frac{\phi_{32}}{2})\right],
\end{aligned}
\end{equation}
\begin{equation}
\label{DeltaU_exact4}
\begin{aligned} 
\Im\big[(\tilde{\mathcal{U}}_{b b} - \tilde{\mathcal{U}}_{a a})\tilde{\mathcal{U}}^*_{a b}\big]& = \frac{(\mathsf{h}_{bb} - \mathsf{h}_{aa})\mathsf{h}_{ab}}{\Omega}\left[\dfrac{\mathscr{D}_{2}\mathscr{C}_{1} - \mathscr{D}_{1}\mathscr{C}_{2}}{\DeltaM{21}} \sin(\phi_{21})\right.\\
& \left.\phantom{=} -\dfrac{\mathscr{D}_{3} \mathscr{C}_{1} - \mathscr{D}_{1} \mathscr{C}_{3}}{\DeltaM{31}}\sin(\phi_{31}) + \dfrac{\mathscr{D}_{3}\mathscr{C}_{2} - \mathscr{D}_{2}\mathscr{C}_{3}}{\DeltaM{32}}\sin(\phi_{32})\right] \\
&= \frac{(\mathsf{h}_{bb} - \mathsf{h}_{aa})\mathsf{h}_{ab}}{\Omega}(q_{4} - q_{3})\Big[\sin(\phi_{21}) - \sin(\phi_{31}) + \sin(\phi_{32})\Big].
\end{aligned}
\end{equation}
From the quantities in Eqs. \eqref{defq123} and \eqref{defq4} it is immediate to see that
\begin{equation}
(\mathsf{h}_{bb} - \mathsf{h}_{aa})\mathsf{h}_{ab}(q_{4} - q_{3}) = \mathsf{h}_{ea}\mathsf{h}_{eb}(q_{1} - q_{2})
\end{equation}
and, therefore, that $\Im\big[(\Utilde_{bb} - \Utilde_{aa})\Utilde^{\ast}_{ab}\big] = \Im(\Utilde_{ea} \, \Utilde^{\ast}_{eb})$, as required by the unitarity condition \eqref{eq2.16}. As a consequence, the CP violation term in the muon-to-tau neutrino conversion probability has the opposite sign to that in the muon-to-electron neutrino conversion probability:
\begin{equation}
\begin{aligned}
\label{P_mu_tau}
\prob{\mu}{\tau} &= \Lambda^{\mu \tau}_{21} \sin[2](\frac{\DeltaM{21}L}{2}) + \Lambda^{\mu \tau}_{31} \sin[2](\frac{\DeltaM{31}L}{2}) 
+ \Lambda^{\mu \tau}_{32} \sin[2](\frac{\DeltaM{32}L}{2})\\
&\phantom{=}+ 8 \dfrac{\Delta_{21} \Delta_{31} \Delta_{32}}{\DeltaM{21} \DeltaM{31} \DeltaM{32}} \mathcal{J} \sin(\frac{\DeltaM{21}L}{2}) \sin(\frac{\DeltaM{31}L}{2}) \sin(\frac{\DeltaM{32}L}{2}),
\end{aligned}
\end{equation}
where
\begin{equation}
\label{LambdaMUTAU31} 
\begin{aligned}
\Lambda^{\mu \tau}_{21} &= \frac{1}{\Omega \DeltaM{21}} \Big[\mtwosin{23}^{2} \Delta_{\mu\tau}^{2} \mathscr{D}_{2} \mathscr{D}_{1} + \msin{13}^2 \mtwosin{12}^{2} (1 - \mtwosin{23}^{2} \mcos{\delta}^{2}) \Delta_{21}^{2} \mathscr{C}_{2} \mathscr{C}_{1} \\
& \phantom{=}- 4 \tfrac{\mtwocos{23}}{\mcos{13}^2} \Delta_{21} \Delta_{\mu\tau} \big(\mathscr{D}_{2} \mathscr{C}_{1} + \mathscr{D}_{1} \mathscr{C}_{2} \big) \mathcal{J}_{r} \mcos{\delta}\Big],\\
\Lambda^{\mu \tau}_{31} & = \frac{-1}{\Omega \DeltaM{31}} \Big[\mtwosin{23}^{2} \Delta_{\mu\tau}^{2} \mathscr{D}_{3} \mathscr{D}_{1} + \msin{13}^{2} \mtwosin{12}^{2} (1 - \mtwosin{23}^{2} \mcos{\delta}^{2}) \Delta_{21}^{2} \mathscr{C}_{3} \mathscr{C}_{1} \\
& \phantom{=}- 4 \tfrac{\mtwocos{23}}{\mcos{13}^2} \Delta_{21} \Delta_{\mu\tau} \big(\mathscr{D}_{3} \mathscr{C}_{1} + \mathscr{D}_{1} \mathscr{C}_{3} \big) \mathcal{J}_{r} \mcos{\delta}\Big], \\
\Lambda^{\mu \tau}_{32} &= \frac{1}{\Omega \DeltaM{32}} \Big[\mtwosin{23}^{2} \Delta_{\mu\tau}^{2} \mathscr{D}_{3} \mathscr{D}_{2} + \msin{13}^{2} \mtwosin{12}^{2} (1 - \mtwosin{23}^{2} \mcos{\delta}^{2}) \Delta_{21}^{2} \mathscr{C}_{3} \mathscr{C}_{2} \\
& \phantom{=}- 4 \tfrac{\mtwocos{23}}{\mcos{13}^2} \Delta_{21} \Delta_{\mu\tau} \big(\mathscr{D}_{3} \mathscr{C}_{2} + \mathscr{D}_{2} \mathscr{C}_{3}\big) \mathcal{J}_{r} \mcos{\delta}\Big]. 
\end{aligned}
\end{equation}

\

The probability of the transformation from $\nu_{\tau}$ to $\nu_{e}$ can be obtained by doing the substitutions 
$\mcos{23} \to - \msin{23}$ and $\msin{23} \to \mcos{23}$ in the formula for $\prob{\mu}{e}$. Proceeding in this way, taking into account the change of the sign of $\mathcal{J}_r$, 
we obtain
\begin{equation}
\label{P_tau_e}
\begin{aligned}
\prob{\tau}{e} &= \Lambda^{\tau e}_{21}\sin[2](\frac{\DeltaM{21}L}{2}) + \Lambda^{\tau e}_{31} \sin[2](\frac{\DeltaM{31}L}{2}) + \Lambda^{\tau e}_{32} \sin[2](\frac{\DeltaM{32}L}{2})\\ 
&\phantom{=}+8\dfrac{\Delta_{21} \Delta_{31} \Delta_{32}}{\DeltaM{21} \DeltaM{31} \DeltaM{32}} \mathcal{J} \sin(\frac{\DeltaM{21}L}{2}) \sin(\frac{\DeltaM{31}L}{2}) 
\sin(\frac{\DeltaM{32}L}{2}),
\end{aligned}
\end{equation}
where 	
\begin{equation}
\label{LambdaTauE}
\begin{aligned}
\Lambda^{\tau e}_{21} &= \dfrac{1}{\Omega \DeltaM{21}} \Big[\msin{23}^2 \mcos{13}^2 \mtwosin{12}^2 \Delta_{21}^2 
\mathscr{A}_{2}\mathscr{A}_{1} + \mcos{23}^2 \mtwosin{13}^2 \Delta_{ee}^2\mathscr{B}_{2} \mathscr{B}_{1}
- 4 \Delta_{21} \Delta_{ee} \big(\mathscr{A}_{2} \mathscr{B}_{1} + \mathscr{B}_{2} \mathscr{A}_{1} \big)\mathcal{J}_{r} \mcos{\delta}\Big],\\
\Lambda^{\tau e}_{31} &= \dfrac{-1}{\Omega \DeltaM{31}} \Big[\msin{23}^2 \mcos{13}^2 \mtwosin{12}^2 \Delta_{21}^2 
\mathscr{A}_{3}\mathscr{A}_{1} + \mcos{23}^2 \mtwosin{13}^2 \Delta_{ee}^2\mathscr{B}_{3}\mathscr{B}_{1}
- 4 \Delta_{21} \Delta_{ee} \big(\mathscr{A}_{3} \mathscr{B}_{1} + \mathscr{B}_{3} \mathscr{A}_{1}\big)\mathcal{J}_{r} \mcos{\delta}\Big],\\
\Lambda^{\tau e}_{32} &= \dfrac{1}{\Omega \DeltaM{32}} \Big[\msin{23}^2 \mcos{13}^2 \mtwosin{12}^2 \Delta_{21}^2 
\mathscr{A}_{3} \mathscr{A}_{2} + \mcos{23}^2 \mtwosin{13}^2 \Delta_{ee}^2\mathscr{B}_{3} \mathscr{B}_{2}
- 4 \Delta_{21} \Delta_{ee} \big(\mathscr{A}_{3} \mathscr{B}_{2} + \mathscr{B}_{3} \mathscr{A}_{2} \big)\mathcal{J}_{r} \mcos{\delta}\Big].
\end{aligned}
\end{equation}

\noindent
The expressions for $\prob{e}{\mu}$, $\prob{\tau}{\mu}$ and $\prob{e}{\tau}$ result immediately from replacing $\delta$ with $-\delta$ in the equations \eqref{P_mu_e}, \eqref{P_mu_tau}, and \eqref{P_tau_e}, respectively. This corresponds to doing $\mathcal{J} \to - \mathcal{J}$ in the CP-violating term, while the expressions for the coefficients $\Lambda^{e \mu}_{kj}, \Lambda^{\tau \mu}_{kj}$, and $\Lambda^{e \tau}_{kj}$ ($kj = 21, 31, 32$) are identical to those for $\Lambda^{\mu e}_{kj}, \Lambda^{\mu \tau}_{kj}$, and $\Lambda^{\tau e}_{kj}$, respectively.

Now consider the survival probabilities $\prob{\alpha}{\alpha}$. The simplest and most direct procedure for obtaining explicit formulas for them consists of applying the conservation of probability with the corresponding flavor-changing probabilities. Accordingly, a fairly compact formula for $\prob{e}{e}$ is deduced from the second equality of Eq. \eqref{Generic_Pee}:
\begin{equation}
\label{Elec_prob3}
\prob{e}{e} = 1 + \Lambda^{e e}_{21} \sin[2](\frac{\DeltaM{21} L}{2}) + \Lambda^{e e}_{31} \sin[2](\frac{\DeltaM{31} L}{2}) + \Lambda^{e e}_{32} \sin[2](\frac{\DeltaM{32} L}{2}),
\end{equation}
where
\begin{equation}
\label{Lambda_Mat_Pee}
\begin{aligned}
\Lambda^{e e}_{21} & = -\frac{1}{\Omega \DeltaM{21}} \Big[\mcos{13}^{2} \mtwosin{12}^{2} \Delta_{21}^{2} 
\mathscr{A}_{2} \mathscr{A}_{1} + \mtwosin{13}^{2} \Delta_{ee}^{2} \mathscr{B}_{2} \mathscr{B}_{1}\Big], \\
\Lambda^{e e}_{31} &= +\frac{1}{\Omega \DeltaM{31}} \Big[\mcos{13}^2 \mtwosin{12}^{2} \Delta_{21}^{2} 
\mathscr{A}_{3} \mathscr{A}_{1} + \mtwosin{13}^{2} \Delta_{ee}^{2} \mathscr{B}_{3} \mathscr{B}_{1}\Big],\\
\Lambda^{e e}_{32}& = -\frac{1}{\Omega \DeltaM{32}} \Big[\mcos{13}^{2} \mtwosin{12}^{2} \Delta_{21}^{2} 
\mathscr{A}_{3} \mathscr{A}_{2} + \mtwosin{13}^{2} \Delta_{ee}^{2} \mathscr{B}_{3} \mathscr{B}_{2}\Big].
\end{aligned} 
\end{equation}
The survival probability of $\nu_{\mu}$ can be written as
\begin{equation}
\label{Muon_mu_mu3}
\begin{aligned}
\prob{\mu}{\mu} &= 1 - \prob{\mu}{e} - \prob{\mu}{\tau} \\
& = 1 + \Lambda^{\mu\mu}_{21} \sin[2](\frac{\DeltaM{21}L}{2}) + \Lambda^{\mu\mu}_{31} \sin[2](\frac{\DeltaM{31}L}{2}) + \Lambda^{\mu\mu}_{32} \sin[2](\frac{\DeltaM{32}L}{2}),
\end{aligned}
\end{equation}
with
\begin{alignat}{3}
\Lambda^{\mu\mu}_{21} &= -(\Lambda^{\mu e}_{21}+\Lambda^{\mu\tau}_{21}), &\qquad
\Lambda^{\mu\mu}_{31} &= -(\Lambda^{\mu e}_{31}+\Lambda^{\mu\tau}_{31}), &\qquad
\Lambda^{\mu\mu}_{32} &= -(\Lambda^{\mu e}_{32}+\Lambda^{\mu\tau}_{32}),
\end{alignat}
where the expressions of the summands are given in Eqs. \eqref{LambdaEMU31} and \eqref{LambdaMUTAU31}. 
A similar formula can also be given for $\prob{\tau}{\tau}$.

All the $\prob{\alpha}{\alpha}$ can be calculated directly by applying the formula in \eqref{Identity1} to the expressions for $\Ucal_{\alpha\alpha}$ given in Eq.~\eqref{DetermForm}, considering the equalities $\omega_{32} - \omega_{31} + \omega_{21} = \omega_{32} \xi_{1} - \omega_{31} \xi_{2} + \omega_{21} \xi_{3} = 0$, and $\omega_{32} \xi_{1}^2 - \omega_{31} \xi_{2}^2 + \omega_{21} \xi_{3}^2 =\Omega$. The results are
\begin{equation}
\label{Surv_prob}
\prob{\alpha}{\alpha} = 1 + \Lambda^{\alpha\alpha}_{21} \sin[2](\frac{\DeltaM{21}L}{2}) + \Lambda^{\alpha\alpha}_{31} \sin[2](\frac{\DeltaM{31}L}{2}) + \Lambda^{\alpha\alpha}_{32} \sin[2](\frac{\DeltaM{32}L}{2}),
\end{equation}
where
\begin{equation}
\label{Prob_ee}
\begin{aligned}
\Lambda^{ee}_{21} &= \phantom{+}\frac{4}{\Omega \DeltaM{21}} 
\begin{vmatrix}
	\mathsf{h}_{\mu\mu} - \xi_{2} & \mathsf{h}^{\ast}_{\mu\tau} \\
	\mathsf{h}_{\mu\tau} & \mathsf{h}_{\tau\tau} - \xi_{2} \\
\end{vmatrix}
\begin{vmatrix}
	\mathsf{h}_{\mu\mu} - \xi_{1} & \mathsf{h}^{\ast}_{\mu\tau} \\
	\mathsf{h}_{\mu\tau} & \mathsf{h}_{\tau\tau} - \xi_{1} \\
\end{vmatrix}, \\
\Lambda^{ee}_{31} &= -\frac{4}{\Omega \DeltaM{31}} 
\begin{vmatrix}
	\mathsf{h}_{\mu\mu} - \xi_{3} & \mathsf{h}^{\ast}_{\mu\tau} \\
	\mathsf{h}_{\mu\tau} & \mathsf{h}_{\tau\tau} - \xi_{3} \\
\end{vmatrix}
\begin{vmatrix}
	\mathsf{h}_{\mu\mu} - \xi_{1} & \mathsf{h}^{\ast}_{\mu\tau} \\
	\mathsf{h}_{\mu\tau} & \mathsf{h}_{\tau\tau} - \xi_{1} \\
\end{vmatrix}, \\
\Lambda^{ee}_{32} &= \phantom{+}\frac{4}{\Omega \DeltaM{32}} 
\begin{vmatrix}
	\mathsf{h}_{\mu\mu} - \xi_{2} & \mathsf{h}^{\ast}_{\mu\tau} \\
	\mathsf{h}_{\mu\tau} & \mathsf{h}_{\tau\tau} - \xi_{2} \\
\end{vmatrix}
\begin{vmatrix}
	\mathsf{h}_{\mu\mu} - \xi_{3} & \mathsf{h}^{\ast}_{\mu\tau} \\
	\mathsf{h}_{\mu\tau} & \mathsf{h}_{\tau\tau} - \xi_{3} \\
\end{vmatrix},
\end{aligned}
\end{equation}

\begin{equation}
\label{Prob_mumu}
\begin{aligned}
\Lambda^{\mu\mu}_{21} &= \phantom{+}\frac{4}{\Omega \DeltaM{21}} 
 \begin{vmatrix}
	\mathsf{h}_{ee} - \xi_{2} & \mathsf{h}^{\ast}_{e\tau} \\
	\mathsf{h}_{e\tau} & \mathsf{h}_{\tau\tau} - \xi_{2} \\
\end{vmatrix}
\begin{vmatrix}
	\mathsf{h}_{ee} - \xi_{1} & \mathsf{h}^{\ast}_{e\tau} \\
	\mathsf{h}_{e\tau} & \mathsf{h}_{\tau\tau} - \xi_{1} \\
\end{vmatrix}, \\
\Lambda^{\mu\mu}_{31} &= -\frac{4}{\Omega \DeltaM{31}} 
\begin{vmatrix}
	\mathsf{h}_{ee} - \xi_{3} & \mathsf{h}^{\ast}_{e\tau} \\
	\mathsf{h}_{e\tau} & \mathsf{h}_{\tau\tau} - \xi_{3} \\
\end{vmatrix} 
\begin{vmatrix}
	\mathsf{h}_{ee} - \xi_{1} & \mathsf{h}^{\ast}_{e\tau} \\
	\mathsf{h}_{e\tau} & \mathsf{h}_{\tau\tau} - \xi_{1} \\
\end{vmatrix}, \\
\Lambda^{\mu\mu}_{32} &= \phantom{+}\frac{4}{\Omega \DeltaM{32}} \begin{vmatrix}
	\mathsf{h}_{ee} - \xi_{2} & \mathsf{h}^{\ast}_{e\tau} \\
	\mathsf{h}_{e\tau} & \mathsf{h}_{\tau\tau} - \xi_{2} \\
\end{vmatrix} 
\begin{vmatrix}
	\mathsf{h}_{ee} - \xi_{3} & \mathsf{h}^{\ast}_{e\tau} \\
	\mathsf{h}_{e\tau} & \mathsf{h}_{\tau\tau} - \xi_{3} \\
\end{vmatrix},
\end{aligned}
\end{equation}

\begin{equation}
\label{Prob_tautau}
\begin{aligned}
\Lambda^{\tau\tau}_{21} &= \phantom{+}\frac{4}{\Omega \DeltaM{21}}
\begin{vmatrix}
	\mathsf{h}_{ee} - \xi_{2} & \mathsf{h}^{\ast}_{e\mu} \\
	\mathsf{h}_{e\mu} & \mathsf{h}_{\mu\mu} - \xi_{2} \\
\end{vmatrix}
\begin{vmatrix}
	\mathsf{h}_{ee} - \xi_{1} & \mathsf{h}^{\ast}_{e\mu} \\
	\mathsf{h}_{e\mu} & \mathsf{h}_{\mu\mu} - \xi_{1} \\
\end{vmatrix}, \\
\Lambda^{\tau\tau}_{31} &= -\frac{4}{\Omega \DeltaM{31}}
\begin{vmatrix}
	\mathsf{h}_{ee} - \xi_{3} & \mathsf{h}^{\ast}_{e\mu} \\
	\mathsf{h}_{e\mu} & \mathsf{h}_{\mu\mu} - \xi_{3} \\
\end{vmatrix} 
\begin{vmatrix}
	\mathsf{h}_{ee} - \xi_{1} & \mathsf{h}^{\ast}_{e\mu} \\
	\mathsf{h}_{e\mu} & \mathsf{h}_{\mu\mu} - \xi_{1} \\
\end{vmatrix}, \\
\Lambda^{\tau\tau}_{32} &= \phantom{+}\frac{4}{\Omega \DeltaM{32}} 
\begin{vmatrix}
	\mathsf{h}_{ee} - \xi_{2} & \mathsf{h}^{\ast}_{e\mu} \\
	\mathsf{h}_{e\mu} & \mathsf{h}_{\mu\mu} - \xi_{2} \\
\end{vmatrix}
\begin{vmatrix}
	\mathsf{h}_{ee} - \xi_{3} & \mathsf{h}^{\ast}_{e\mu} \\
	\mathsf{h}_{e\mu} & \mathsf{h}_{\mu\mu} - \xi_{3} \\
\end{vmatrix}.
\end{aligned}
\end{equation}
Equations \eqref{Surv_prob}--\eqref{Prob_tautau} follow directly from the Lagrange-Sylvester formula for the flavor-basis Hamiltonian and therefore remain valid for any constant Hermitian Hamiltonian, e.g., when non-standard interactions are included, in which case the transformation of Eq.~\eqref{Ham_matter2} does not yield a Hamiltonian independent of $\theta_{23}$ and $\delta$. In the standard case, the equality in \eqref{detequality} shows that the determinants in Eq.~\eqref{Prob_ee} can be evaluated in the intermediate basis, which makes explicit that $\prob{e}{e}$ does not depend on $\theta_{23}$ and $\delta$. However, for the situation analyzed in this work, where $V$ is given by the standard charged-current interaction of neutrinos in ordinary matter, the most efficient procedure for deriving explicit formulas for all $\prob{\alpha}{\alpha}$ is to apply probability conservation with the appropriate flavor-change probabilities.

In summary, the transition probabilities in matter can be expressed by the formula
\begin{equation}
\label{Global3}
\prob{\alpha}{\beta} =\delta_{\alpha \beta} + \sum\limits_{j<k} \Lambda^{\alpha \beta}_{kj} \sin^{2}{\left(\dfrac{\DeltaM{kj} L}{2}\right)} 
+ \epsilon_{\alpha\beta\gamma} 8 \mathcal{J} \prod\limits_{j<k} \frac{\Delta_{kj}}{\DeltaM{kj}} \sin(\dfrac{\DeltaM{kj} L}{2}),
\end{equation}
where the coefficients $\Lambda^{\alpha \beta}_{kj}$ are equivalent to the quantities $-4 \Re(U^{m\ast}_{\alpha k} U^{m}_{\beta k} U^{m}_{\alpha j} U^{m\ast}_{\beta j})$, with $U^{m}_{\alpha j}$ elements of the mixing matrix in matter. In this way, we have derived an exact analytical solution to the problem of neutrino oscillations in matter of constant density and presented the resulting flavor transition probabilities in a form analogous to the vacuum expressions. As shown in Appendix~\ref{AppB}, in the limit of zero density these probabilities reduce to  the vacuum expressions presented in Section~\ref{Section2}.  Although the calculations are too long to present here, it is worth mentioning that the expressions derived by expanding the exact probabilities to the first order in $\msin{13}$ or in the mass hierarchy correspond exactly to those obtained in Ref.~\cite{Akhmedov2004} from perturbative solutions of the flavor evolution matrix.
We also note that the coefficients $\Lambda^{\alpha\beta}_{31}$ and $\Lambda^{\alpha\beta}_{32}$ in all channels contain terms proportional to $\Delta_{21}^{2}$, which, in turn, are multiplied by the factors $\mathscr{A}_{3}$ or $\mathscr{C}_{3}$, as appropriate. All these terms are of order $\alpha^{2}$, where $\alpha = \Delta_{21}/\Delta_{ee} \approx 0.03$, and vanish when the potential is set to zero. Their relevance is discussed in Section~\ref{Conclusions}.

\section{Hybrid approach}
\label{hybridsec}

The exact eigenvalues of the neutrino energies in matter are the only ingredient that prevents the oscillation probabilities presented in the previous section from being completely transparent. The formulas in Eq.~\eqref{eigenvalues} have a geometric origin \cite{Nickalls:2006} and contain the cosine of an inverse cosine; this obscures the dependence of the eigenvalues on the oscillation parameters and the potential, making them impractical for phenomenological analyses. However, one should note that, apart from the eigenvalues $ \xi_{k}$, the quantities in Eqs. \eqref{defA,B} and \eqref{defC,D} depend on the oscillation parameters and the potential only through the matrix entries of $\tilde{H}$. Consequently, the forms of the coefficients 
$\Lambda^{\alpha\beta}_{kj}$ in Eq. \eqref{Global3} do not depend on the explicit expressions of the $ \xi_{k}$. Therefore, we can replace the exact eigenvalues with approximate ones without altering the structure of the probabilities. We will designate this procedure as the hybrid method.

Several sets of approximate eigenvalues are available in the literature, determined by perturbation theory or by successive rotations of the Hamiltonian \cite{Agarwalla2014, Minakata2016, Denton2016}. Some of these exhibit level crossings and are accurate only far from the corresponding resonances, a fact that must be considered when selecting them for a hybrid approach. Here, we use eigenvalues obtained from two successive rotations of  $\tilde{H}$, adopting---with some modifications---the scheme presented in Ref.~\cite{Denton2016}: a rotation in the 1--3 plane, which removes off-diagonal entries proportional to $V$, followed by a rotation in the 1--2 plane, which takes into account the solar resonance. Details of their derivation are given in Appendix~\ref{AppC}. The resulting eigenvalues are:
\begin{equation}
\label{ApproxEigen}
\begin{aligned}
\zeta_{1} &= \dfrac{1}{2} \left(\Delta_{21} - \Delta_{l} + \hat{V}\right), \\
\zeta_{2} &= \dfrac{1}{2} \left(\Delta_{21} + \Delta_{l} + \hat{V}\right), \\
\zeta_{3} &= \dfrac{1}{2} \left(\msin{12}^{2} \Delta_{21} + \Delta_{31} + V + \varepsilon\Delta_{h}\right),
\end{aligned}
\end{equation}
where $\varepsilon \equiv \mathrm{sgn}(\Delta_{ee})$,
\begin{equation}
\label{ApproxEigenDefs}
\Delta_{h} = \sqrt{(\mtwocos{13} \Delta_{ee} - V)^{2} + \mtwosin{13}^{2} \Delta_{ee}^{2}}, \qquad
\Delta_{l} = \sqrt{(\mtwocos{12} \Delta_{21} - \hat{V})^{2} + \mcos{\eta}^{2}\, \mtwosin{12}^{2} \Delta_{21}^{2}},
\end{equation}
and
\begin{equation}
\label{VhatCeta}
\hat{V} = \tfrac{1}{2}\left(\Delta_{ee} + V - \varepsilon\Delta_{h}\right), \qquad
\mcos{\eta}^{2} = \frac{\Delta_{h} + \varepsilon\left(\Delta_{ee} - \mtwocos{13} V\right)}{2\Delta_{h}}.
\end{equation}
Here, $\hat{V}$ plays the role of an effective potential in the 1--2 sector, and $\eta$ is the angle of the 1--3 rotation, which enters only through $\mcos{\eta}^{2}$ and is written above without reference to any angle. For antineutrinos, the eigenvalues are obtained by replacing $V \to -V$. The rotation angles are auxiliary quantities in the derivation and do not appear in the hybrid probabilities; hence, the matter dependence of the latter is entirely contained in the eigenvalues and in the entries of $\tilde{H}$. The off-diagonal entries neglected when obtaining Eq.~\eqref{ApproxEigen} are proportional to $\mtwosin{12}\msin{\eta}\Delta_{21}$ [see Eq.~\eqref{RotHfinal}], and $\msin{\eta}$ grows as the atmospheric resonance is approached.

The eigenvalues in Eq. \eqref{ApproxEigen} have two properties that make them particularly suitable for the hybrid approach. First, their sum is exact, $\zeta_{1} + \zeta_{2} + \zeta_{3} = \Delta_{21} + \Delta_{31} + V = {\Tr}(\tilde{H})$, so the first of Vieta's formulas in \eqref{vietaformulas} is preserved. Second, for $V = 0$ we have $\eta = 0$, $\hat{V} = 0$, $\varepsilon\Delta_{h} = \Delta_{ee}$, and $\Delta_{l} = \Delta_{21}$, so that $\zeta_{1} = 0$, $\zeta_{2} = \Delta_{21}$, and $\zeta_{3} = \Delta_{31}$. Consequently, in this limit, just like the exact probabilities, the hybrid probabilities reproduce the exact expressions in vacuum.

The hybrid expressions for $P_{\mu e}$, $P_{\mu\tau}$, and $P_{ee}$ are obtained by substituting $\xi_{k}$ with $ \zeta_{k}$ in the respective coefficients given in Eqs. \eqref{LambdaEMU31}, \eqref{LambdaMUTAU31}, and \eqref{Lambda_Mat_Pee}, also taking into account that  $\DeltaM{kj} = \zeta_{k} - \zeta_{j}$. Note that Eq.~\eqref{Trifactor} must not be used in this case, since it gives $\Omega$ in terms of the exact invariants $\mathsf{a}$, $\mathsf{b}$, and $\mathsf{c}$. The resulting probabilities are exactly unitary for any choice of approximate eigenvalues: since $\Lambda^{ee}_{kj} = -(\Lambda^{\mu e}_{kj} + \Lambda^{\tau e}_{kj})$ holds identically and the CP-violating terms of all channels share the same factor, the conditions $\sum_{\beta} \prob{\alpha}{\beta} = \sum_{\alpha} \prob{\alpha}{\beta} = 1$ are satisfied exactly,  assuming that the exact survival probabilities $\prob{\mu}{\mu}$ and $\prob{\tau}{\tau}$ were obtained using probability conservation. Thus, unlike in standard perturbation theory, unitarity does not have to be imposed order by order.

\begin{figure}[!ht]
\centering
\begin{subfigure}[b]{0.5\textwidth}
\includegraphics[width=\textwidth]{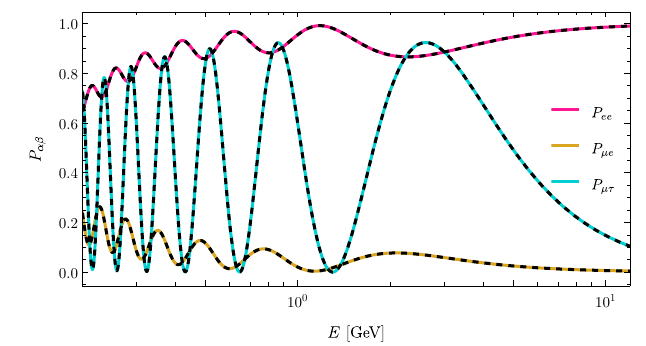}
\caption{Neutrino oscillation probabilities}
\label{fig2a}
\end{subfigure}\begin{subfigure}[b]{0.5\textwidth}
\includegraphics[width=\textwidth]{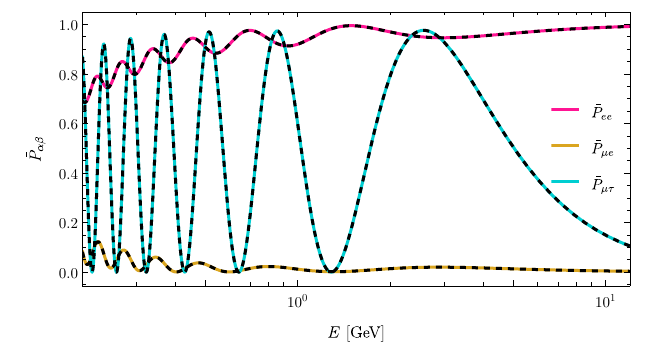}
\caption{Antineutrino oscillation probabilities}
\label{fig2b}
\end{subfigure}
\hfill
\begin{subfigure}[b]{0.5\textwidth}
\centering
\includegraphics[width=\textwidth]{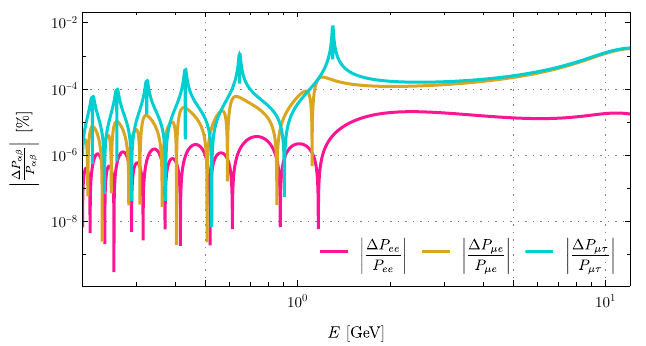}
\caption{Relative error in neutrino probabilities}
\label{fig2c}
\end{subfigure}\begin{subfigure}[b]{0.5\textwidth}
\includegraphics[width=\textwidth]{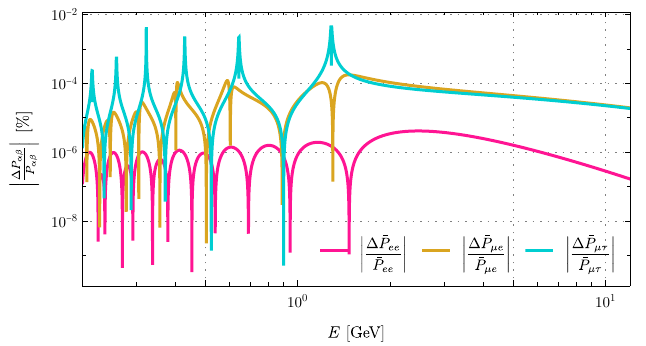}
\caption{Relative error in antineutrino probabilities}
\label{fig2d}
\end{subfigure}
\caption{Oscillation probabilities $P_{ee}$, $P_{\mu e}$, and $P_{\mu\tau}$ for a DUNE-like configuration ($L = 1300$~km, $\rho = 3~\mathrm{g/cm^{3}}$, normal ordering, NuFit-6.0 best-fit values \cite{Esteban:2024eli}). (a) and (b): exact results (solid colored lines) and hybrid results obtained with the eigenvalues \eqref{ApproxEigen} (black dashed lines), for neutrinos and antineutrinos, respectively. (c) and (d): relative percentage error of the hybrid probabilities. The sharp peaks correspond to minima of the probabilities, where the denominator of the relative error becomes very small.}
\label{Figure2}
\end{figure}

To evaluate the performance of the hybrid approach, we consider a configuration similar to that of DUNE: a baseline $L = 1300$~km, a constant matter density $\rho = 3~\mathrm{g/cm^{3}}$, with electron fraction of $0.4957$, normal ordering, and the best-fit values NuFit-6.0 (IC24 with SK atmospheric data) \cite{Esteban:2024eli}, namely, $\sin^{2}\theta_{12} = 0.308$, $\sin^{2}\theta_{13} = 0.02215$, $\sin^{2}\theta_{23} = 0.470$, $\delta = 212^{\circ}$, $\Delta m^{2}_{21} = 7.49 \times 10^{-5}~\mathrm{eV^{2}}$, and $\Delta m^{2}_{31} = 2.513 \times 10^{-3}~\mathrm{eV^{2}}$. In Figures~\ref{fig2a} and \ref{fig2b}, we plot $P_{ee}$, $P_{\mu e}$, and $P_{\mu\tau}$ for neutrinos and antineutrinos, respectively, in the energy range of $0.2 \leq E \leq 12$~GeV. As can be seen, the exact and hybrid curves are indistinguishable at the scale of the figures. Figures~\ref{fig2c} and \ref{fig2d} show the relative percentage error of the hybrid probabilities. In these three channels, for both neutrinos and antineutrinos, it remains below $0.01 \%$ across the entire energy range, and the absolute errors are at most about $2 \times 10^{-6}$.  In other channels, such as $P_{\mu\mu}$ and $P_{\tau\tau}$, the relative error can exceed $0.01\%$ close to their minima, where the probabilities themselves are small, while the absolute errors remain at the same level. Among the channels shown, the largest relative errors, about $8 \times 10^{-3}\%$ for neutrinos and $5 \times 10^{-3}\%$ for antineutrinos, occur in $P_{\mu\tau}$ close to its minima, where the probability itself is below $10^{-4}$; they reflect the smallness of the denominator rather than a deterioration of the approximation. For $P_{ee}$, the relative error does not exceed about $2 \times 10^{-5}\%$. For $P_{\mu e}$, the channel most relevant to CP-violation searches, it stays below $3 \times 10^{-4}\%$ for $0.5 \leq E \leq 5$~GeV, where the DUNE flux is concentrated, and grows slowly toward higher energies, reaching about $1.7 \times 10^{-3}\%$ at $12$~GeV as the atmospheric resonance ($E \approx 10$~GeV for this density) is approached. For antineutrinos, which do not undergo this resonance in normal ordering, the relative error in $\bar{P}_{\mu e}$ stays below $2 \times 10^{-4}\%$ over the whole range.

\begin{figure}[!ht]
\centering
\begin{subfigure}[b]{0.5\textwidth}
\includegraphics[width=\textwidth]{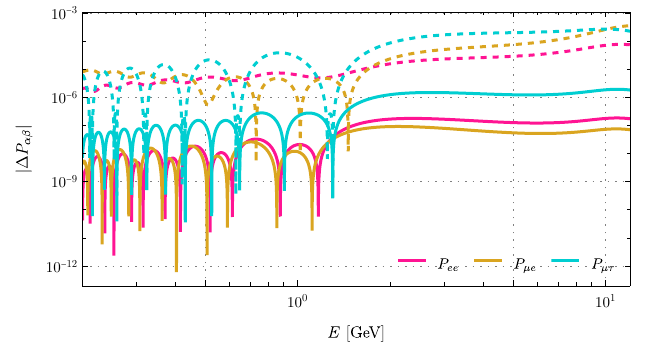}
\caption{Neutrinos}
\label{fig3a}
\end{subfigure}\begin{subfigure}[b]{0.5\textwidth}
\includegraphics[width=\textwidth]{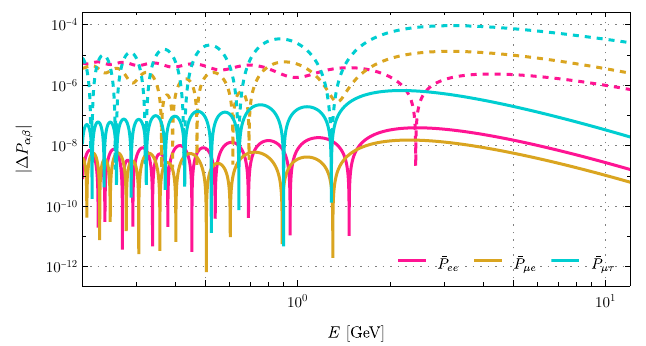}
\caption{Antineutrinos}
\label{fig3b}
\end{subfigure}
\caption{Absolute errors of the hybrid probabilities (solid lines) and of the zeroth-order probabilities of the two-rotation method of Ref.~\cite{Denton2016} (dashed lines), for the same configuration as in Figure~\ref{Figure2}. Both approximations use the eigenvalues \eqref{ApproxEigen}.}
\label{Figure3}
\end{figure}

It is instructive to compare these results with the zeroth-order two-rotation probabilities of Ref.~\cite{Denton2016}, which use the same eigenvalues \eqref{ApproxEigen} together with the mixing angles in matter $\theta^{m}_{13}$ and $\theta^{m}_{12}$ (see Appendix~\ref{AppC}). In the same configuration, the maximum absolute errors of the latter are $7.5 \times 10^{-5}$, $3.5 \times 10^{-4}$, and $2.6 \times 10^{-4}$ in $P_{ee}$, $P_{\mu e}$, and $P_{\mu\tau}$, respectively, and $5.8 \times 10^{-6}$, $1.3 \times 10^{-5}$, and $9.2 \times 10^{-5}$ for antineutrinos (see Figure~\ref{Figure3}). These values are larger than those of the hybrid formulas by factors between about $10^{2}$ and $4 \times 10^{3}$, the largest difference occurring in $P_{\mu e}$, whose relative error with the two-rotation formulas grows with energy, reaching $0.3\%$ at $5$~GeV and about $8\%$ at $12$~GeV. Since both approaches share the same eigenvalues, this improvement comes entirely from the coefficients: the hybrid formulas keep the exact entries of $\tilde{H}$, including the terms of order $\alpha^{2}$ discussed in Section~\ref{ProbaSection}, whereas the zeroth-order two-rotation probabilities neglect the residual entries of Eq.~\eqref{RotHfinal}. These entries are proportional to $\msin{\eta}$, which grows as the atmospheric resonance is approached; in the present configuration, $\msin{\eta} \approx 0.05$ at $3$~GeV and $0.5$ at $10$~GeV.

These results show that the hybrid formulas reproduce the exact probabilities to an accuracy far beyond any foreseeable experimental sensitivity, while keeping all matter terms explicit in the coefficients $\Lambda^{\alpha\beta}_{kj}$, which makes them a convenient tool for theoretical analyses.

\section{Discussion and conclusions}
\label{Conclusions}
Neutrino propagation in a medium requires an understanding of the effects of density, and in some cases, compositional variations. In situations  as in the Earth, the Sun, and supernovae, where the effective potential energy of neutrinos in the medium can be approximated as spherically symmetric, the probabilities of flavor transformations have regularity properties that facilitate their calculation. In this work, we have derived explicit expressions for three-neutrino oscillations using the evolution operator formalism in matter of constant density. Applying Lagrange-Sylvester interpolation and Vieta's formulas, we have provided compact and exact expressions for the components of the evolution operator in the intermediate basis. This, in turn, allowed us to derive the oscillation probabilities  in terms of coefficients $\Lambda^{\alpha \beta}_{kj}$, equivalent to $-4\,\mathrm{Re}(U^{m\ast}_{\alpha k} U^{m}_{\beta k} U^{m}_{\alpha j} U^{m\ast}_{\beta j})$, written in symmetric and compact form in terms of the oscillation parameters and the eigenvalues of the Hamiltonian.

The expressions for $P_{ee}$, $P_{\mu e}$, and $P_{\mu \tau}$ given by Eqs.~\eqref{Elec_prob3}, \eqref{P_mu_e}, and \eqref{P_mu_tau}, together with their coefficients, are the central results of this work. They do not involve mixing angles in matter, making the dependence on the CP-violating phase transparent. Genuine CP violation appears in a single, matter dependent term, proportional to $\mathcal{J}$, with the remaining dependence on $\delta$ entering in the coefficients $\Lambda^{\mu e, \,\mu \tau}_{kj}$ only through $\cos\delta$. In $\Lambda^{\mu e}_{kj}$, it appears as a single term proportional to $\mathcal{J}_{r}\cos\delta$, whose coefficient depends on the matter potential through the eigenvalues. This coefficient is different for neutrinos and antineutrinos and does not cancel in $P_{\mu e} - \bar{P}_{\mu e}$. Consequently, the difference $[P_{\mu e}(\delta) - \bar{P}_{\mu e}(\delta)] - [P_{\mu e}(0) - \bar{P}_{\mu e}(0)]$, sometimes used to remove matter-induced asymmetry, does not vanish for $\delta = \pi$, where CP is conserved: the $\cos\delta$ terms contribute $(\cos\delta - 1)$ times an odd coefficient in $V$ and thus vanish in vacuum.
Furthermore, the exact coefficients  $\Lambda^{\alpha\beta}_{31}$ and $\Lambda^{\alpha\beta}_{32}$ contain terms of order $\alpha^{2}$ that vanish for $V = 0$. Most perturbative expansions, including the zeroth-order approximation using double rotation, do not capture these terms, which, although very small, become significant near and beyond the atmospheric resonance. Our numerical comparison clarifies this, as discussed below.

The main drawback of using exact solutions in phenomenological analyses lies in the formulas for exact energy eigenvalues, given by trigonometric expressions that obscure their dependence on oscillation parameters and the potential in matter. Since the structure of our exact probabilities for flavor transitions does not depend on explicit eigenvalue expressions, we implement a hybrid approach in which these probabilities are evaluated in terms of approximate eigenvalues. These eigenvalues are expressed in terms of radicals and were obtained following the strategy of Ref.~\cite{Denton2016}, by means of a pair of successive rotations in sectors 1--3 and 1--2 of the matrix $\tilde{H}$. The sum of the eigenvalues thus derived is equal to the trace of  $\tilde{H}$, as required by one of Vieta's identities, and they reduce to the vacuum values for $V = 0$. The eigenvalues of the antineutrinos result simply from substituting $V \to - V$ into the neutrino expressions.

Our procedure for calculating the evolution matrix requires only the eigenvalues, not the matrix used to diagonalize the Hamiltonian. Therefore, the hybrid probabilities, like the exact ones, do not involve the mixing angles or the CP violation phase in matter. For a DUNE-like configuration, the hybrid probabilities reproduce the exact probability values with absolute errors of about $2 \times 10^{-6}$ in all channels and relative errors less than $0.01 \%$ for $P_{ee}$, $P_{\mu e}$, and $P_{\mu\tau}$. In these channels, the absolute errors are at least two orders of magnitude smaller than those of the zeroth-order two-rotation probabilities, which use the same eigenvalues. Since both procedures share the same eigenvalues, the improvement we obtained is due to the aforementioned  $\alpha^{2}$ order terms in the exact coefficients, which are maintained in the hybrid approach and become important near the atmospheric resonance. The accuracy of the hybrid approach is limited only by that of the eigenvalues and can therefore be systematically improved by using better eigenvalues. Furthermore, the survival probabilities expressed in terms of the principal minors of the Hamiltonian in the flavor basis, Eqs.~\eqref{Prob_ee}--\eqref{Prob_tautau}, do not depend on the transformation of Eq.~\eqref{Ham_matter2} and can even be applied when non-standard interactions are included.

\acknowledgments
This work was supported by CONAHCYT through fellowship 823808 and projects CB-240666 and CF-2023-1169, as well as by DGAPA-UNAM through project PAPIIT-IN108917 and the PASPA fellowship.

\appendix
\section{Technical details}
\label{AppA}
Here, we give the generic formulas needed to calculate the modulus squared $|\Utilde_{\ell k}|^2$ ($|\Ucal_{\ell k}|^2$) and the real and imaginary parts of the products $\Utilde_{\ell k} \, \Utilde^{\ast}_{\ell m}$ ($\Ucal_{\ell k} \, \Ucal^{\ast}_{\ell m}$) of the matrix elements of the evolution operator in the intermediate (flavor) basis. 
\begin{equation}
\label{Identity1}
\begin{aligned}
\abs{A_1 - A_2 e^{-i \phi_{21}} + A_3 e^{-i \phi_{31}}}^2 & = (A_1 - A_2 + A_3)^2 + 4 A_1 A_2 \sin[2](\frac{\phi_{21}}{2}) \\
\phantom{=} & - 4 A_1 A_3 \sin[2](\frac{\phi_{31}}{2})+ 4 A_2 A_3 \sin[2](\frac{\phi_{32}}{2}),
\end{aligned}
\end{equation}
\begin{equation}
\label{Identity2} 
\begin{aligned}
\Re \big[(A_1 -& A_2 e^{-i \phi_{21}}+ A_3 e^{-i \phi_{31}})(B_1 - B_2 e^{i \phi_{21}} + B_3 e^{i \phi_{31}})\big] \\
&= (A_1 - A_2 + A_3)(B_1 - B_2 + B_3) + 2(A_1 B_2 + A_2 B_1) \sin[2](\frac{\phi_{21}}{2}) \\
\phantom{=}&-2(A_1 B_3 + A_3 B_1) \sin[2](\frac{\phi_{31}}{2})+ 2(A_2 B_3 + A_3 B_2) \sin[2](\frac{\phi_{32}}{2}),
\end{aligned}
\end{equation}
\begin{equation}
\label{Identity3}
\begin{aligned}
\Im\big[(A_1 - A_2 e^{-i \phi_{21}}&+ A_3 e^{-i \phi_{31}})(B_1 - B_2 e^{i \phi_{21}} + B_3 e^{i \phi_{31}})\big] \\
&= (A_2 B_1 - A_1 B_2) \sin{\phi_{21}} + (A_1 B_3 - A_3 B_1) \sin{\phi_{31}}\\ 
&\phantom{=}+ (A_3 B_2 - A_2 B_3) \sin{\phi_{32}},
\end{aligned}
\end{equation}
where $A_{s}$ and $B_{s}$, $s = 1, 2, 3$, are real quantities. Note that for $A_{s} = B_{s}$ the imaginary part vanishes and the real part reduces to \eqref{Identity1}, as it should. 

\section{Limit $V\to 0$}
\label{AppB}

First, let us note that, in this limit, $\Omega = \Delta_{21} \Delta_{31} \Delta_{32}$ and the quantities in Eq.~\eqref{defA,B} become
\begin{alignat}{3}
\label{reduction1}
\mathscr{A}_{1} &= -\Delta_{31}, &\qquad
\mathscr{A}_{2} &= -\Delta_{32}, &\qquad
\mathscr{A}_{3} &= 0,
\end{alignat}
\begin{alignat}{3}
\label{reduction2}
\mathscr{B}_{1} &= -\mcos{12}^{2} \dfrac{\Delta_{21} \Delta_{31}}{\Delta_{ee}}, &\qquad
\mathscr{B}_{2} &= \msin{12}^{2}\dfrac{\Delta_{21} \Delta_{32}}{\Delta_{ee}}, &\quad\text{and }\quad
\mathscr{B}_{3} &= \dfrac{\Delta_{31} \Delta_{32}}{\Delta_{ee}}.
\end{alignat}
Replacing them into the coefficients in \eqref{Lambda_Mat_Pee}, using $\Delta_{ee} - \mcos{12}^{2} \Delta_{31} = \msin{12}^{2} \Delta_{32}$, $\Delta_{ee} - \mcos{12}^{2} \Delta_{21} = \Delta_{32}$, and $\Delta_{31} - \Delta_{21} = \Delta_{32}$, we get
\begin{equation}
\label{Lambda_Mat_Pee3}
\begin{aligned}
\Lambda^{e e}_{21} &= -\mcos{13}^{2} \mtwosin{12}^{2} + \mtwosin{13}^{2} \msin{12}^{2} \mcos{12}^{2} = -\mcos{13}^{4} \mtwosin{12}^{2} ,\\
\Lambda^{e e}_{31} &= -\msin{2\theta_{13}}^2 \mcos{12}^{2},\\
\Lambda^{e e}_{32} &= -\msin{2\theta_{13}}^2 \msin{12}^{2},
\end{aligned}
\end{equation}
which coincides with the expressions given in \eqref{Lambda_Vac_Pee}.

For the conversion probability $\prob{\mu}{e}$, substituting $\mathscr{A}_{i}$ and $\mathscr{B}_{i}$ as given by \eqref{reduction1} and \eqref{reduction2} into the expressions for $\Lambda^{\mu e}_{kj}$ in Eq.~\eqref{LambdaEMU31}, we immediately obtain 
\begin{equation}
\label{LambdamueV0}
\begin{aligned}
\Lambda^{\mu e}_{21} &= \mcos{23}^2 \mcos{13}^2 \mtwosin{12}^2 - \frac{1}{4} \msin{23}^2 \msin{2\theta_{13}}^2 \msin{2\theta_{12}}^2 + 4 \mtwocos{12} \mathcal{J}_{r} \mcos{\delta}, \\
\Lambda^{\mu e}_{31} &= \msin{23}^2 \mtwosin{13}^2 \mcos{12}^{2} + 4 \mathcal{J}_{r} \mcos{\delta},\\
\Lambda^{\mu e}_{32} &= \msin{23}^2 \mtwosin{13}^2 \msin{12}^{2} - 4 \mathcal{J}_{r} \mcos{\delta}.
\end{aligned}
\end{equation}

To examine the limit of the coefficients in $\prob{\mu}{\tau}$, we note that
\begin{alignat}{3}
\label{reduction3}
\mathscr{C}_{1} &= -\Delta_{31}, &\qquad
\mathscr{C}_{2} &= -\Delta_{32}, &\qquad
\mathscr{C}_{3} &= 0,
\end{alignat}
and that the quantities $\Delta_{\mu\tau}\mathscr{D}_{i}$, with $\Delta_{\mu\tau}$ given in \eqref{defD}, after a little algebra, reduce to
\begin{equation}
\label{reduction4}
\begin{aligned}
\Delta_{\mu\tau}\mathscr{D}_{1} &= -(\msin{12}^{2} - \msin{13}^{2} \mcos{12}^{2}) \Delta_{21} \Delta_{31}, \\
\Delta_{\mu\tau}\mathscr{D}_{2} &= -(\msin{13}^{2} \msin{12}^{2} - \mcos{12}^{2}) \Delta_{21} \Delta_{32}, \\
\Delta_{\mu\tau}\mathscr{D}_{3} &= \mcos{13}^{2} \Delta_{31} \Delta_{32}.	
\end{aligned}
\end{equation}
Replacing the corresponding quantities in $\Lambda^{\mu \tau}_{kj}$ (Eq.~\eqref{LambdaMUTAU31}) with their expressions in \eqref{reduction3} and \eqref{reduction4}, 
we find   
\begin{equation}
\label{LambdamutauV0}
\begin{aligned}
\Lambda^{\mu \tau}_{21} &= \mtwosin{23}^{2} (\msin{13}^{2} \msin{12}^{2} 
- \mcos{12}^{2})(\msin{12}^{2} - \msin{13}^{2} \mcos{12}^{2})\\ 
\phantom{=}&+ \msin{13}^{2} \mtwosin{12}^{2} (1 - \mtwosin{23}^{2} \mcos{\delta}^{2}) - 4 \tfrac{\mtwocos{23}}{\mcos{13}^{2}}  \left[(\msin{13}^{2} \msin{12}^{2} - \mcos{12}^{2})+ (\msin{12}^{2} - \msin{13}^{2} \mcos{12}^{2})\right] \mathcal{J}_{r} \mcos{\delta}, \\
&= \mtwosin{23}^{2} (\msin{13}^{2} \msin{12}^{2} - \mcos{12}^{2})(\msin{12}^{2} - \msin{13}^{2} \mcos{12}^{2}) + (1 - \mtwosin{23}^{2} \mcos{\delta}^{2}) \msin{13}^{2} \msin{2\theta_{12}}^{2} +4 \mtwocos{23} \left(2\mcos{13}^{-2} - 1\right) \mcos{2\theta_{12}} \mathcal{J}_{r} \mcos{\delta}, \\
\Lambda^{\mu \tau}_{31} &= \mtwosin{23}^2 \mcos{13}^2 (\msin{12}^2 - \msin{13}^2 \mcos{12}^2) - 4\mtwocos{23}\mathcal{J}_{r} \mcos{\delta}, \\
\Lambda^{\mu \tau}_{32} &= \mtwosin{23}^2 \mcos{13}^2 (\mcos{12}^2 - \msin{13}^2 \msin{12}^2) + 4\mtwocos{23}\mathcal{J}_{r} \mcos{\delta}.	
\end{aligned}
\end{equation}
The expressions for $\Lambda^{\mu e}_{kj}$ and $\Lambda^{\mu \tau}_{kj}$ given in Eqs. \eqref{Lambda_vac_mue} and \eqref{Lambda_vac_mutau} are identical to those in Eqs.~\eqref{LambdamueV0} and \eqref{LambdamutauV0}, respectively. In this way, we have completed the proof that our expressions for the oscillation probabilities in matter reduce to those in vacuum when the potential equals zero.

\section{Approximate eigenvalues via two rotations}
\label{AppC}

Here we derive the eigenvalues \eqref{ApproxEigen} following the two-rotation strategy of Ref.~\cite{Denton2016}. The procedure consists of two successive rotations of $\tilde{H}$, each of which removes one off-diagonal entry exactly. The rotations themselves involve no approximation; the only approximation consists in neglecting, at the end, the off-diagonal entries that remain, which are suppressed by $\alpha = \Delta_{21}/\Delta_{ee} \approx 0.03$.
 
We first undo the vacuum 1--3 rotation. Since $\tilde{H} = \rotmat{13}\rotmat{12} H_{0} \rotmat{12}^{\mathtt{t}} \rotmat{13}^{\mathtt{t}} + VY$, we have $\rotmat{13}^{\mathtt{t}} \tilde{H} \rotmat{13} = \rotmat{12} H_{0} \rotmat{12}^{\mathtt{t}} + V\rotmat{13}^{\mathtt{t}} Y \rotmat{13}$, where the first term contains only the vacuum 1--2 mixing and the second one distributes the potential over the 1--3 block:
\begin{equation}
\label{RotH13}
\rotmat{13}^{\mathtt{t}} \tilde{H} \rotmat{13} = \begin{pmatrix}
\msin{12}^{2} \Delta_{21} + \mcos{13}^{2} V & \tfrac{1}{2} \mtwosin{12} \Delta_{21} & \tfrac{1}{2} \mtwosin{13} V \\[3pt]
\tfrac{1}{2} \mtwosin{12} \Delta_{21} & \mcos{12}^{2} \Delta_{21} & 0 \\[3pt]
\tfrac{1}{2} \mtwosin{13} V & 0 & \Delta_{31} + \msin{13}^{2} V
\end{pmatrix}.
\end{equation}
The 2--3 entry vanishes exactly, the 1--2 entries are of order $\alpha\Delta_{ee}$, and the 1--3 entry is proportional to $V$. Depending on the signs of $V$ and $\Delta_{ee}$, the 1--1 entry equals the 3--3 entry for $\mtwocos{13} V = \Delta_{ee}$ and the 2--2 entry for $\mcos{13}^{2} V = \mtwocos{12}\Delta_{21}$, that is, close to the atmospheric and solar resonances, respectively. Near these points, the off-diagonal entry that couples the corresponding levels is not small compared with the difference of their diagonal entries and must be removed exactly.
 
Leaving aside for the moment the 1--2 entries, the 1--3 block of Eq.~\eqref{RotH13}, apart from the term $\msin{12}^{2}\Delta_{21}$ on its diagonal, is the two-flavor Hamiltonian in matter for $\Delta_{ee} = \Delta_{31} - \msin{12}^{2}\Delta_{21}$ and $\theta_{13}$, written in the vacuum mass basis. It is diagonalized by a rotation $\rotmat{\eta}$ in the 1--3 plane, whose angle is fixed by requiring that the 1--3 entry vanish:
\begin{equation}
\tan 2\eta = \frac{\mtwosin{13} V}{\Delta_{ee} - \mtwocos{13} V}.
\end{equation}
Using $(\Delta_{ee} - \mtwocos{13} V)^{2} + \mtwosin{13}^{2} V^{2} = (\mtwocos{13} \Delta_{ee} - V)^{2} + \mtwosin{13}^{2} \Delta_{ee}^{2} = \Delta_{h}^{2}$, with $\Delta_{h}$ given in Eq.~\eqref{ApproxEigenDefs}, we get
\begin{equation}
\label{etadef}
\cos 2\eta = \frac{\varepsilon\left(\Delta_{ee} - \mtwocos{13} V\right)}{\Delta_{h}}, \qquad \sin 2\eta = \frac{\varepsilon\,\mtwosin{13} V}{\Delta_{h}},
\end{equation}
where the factor $\varepsilon$ selects the solution with $\eta \to 0$ as $V \to 0$ for both mass orderings; the expression for $\mcos{\eta}^{2}$ in Eq.~\eqref{VhatCeta} follows from $\mcos{\eta}^{2} = \tfrac{1}{2}(1 + \cos 2\eta)$. This rotation leaves the 2--2 entry unchanged and replaces the diagonal entries of the 1--3 block by the eigenvalues of that block, so that the diagonal entries become
\begin{equation}
\label{lambdas}
\begin{aligned}
\lambda_{1} &= \tfrac{1}{2}\left(\msin{12}^{2}\Delta_{21} + \Delta_{31} + V - \varepsilon\Delta_{h}\right) = \msin{12}^{2}\Delta_{21} + \hat{V}, \\
\lambda_{2} &= \mcos{12}^{2}\Delta_{21}, \\
\lambda_{3} &= \tfrac{1}{2}\left(\msin{12}^{2}\Delta_{21} + \Delta_{31} + V + \varepsilon\Delta_{h}\right) = \Delta_{31} + V - \hat{V},
\end{aligned}
\end{equation}
with $\hat{V}$ given in Eq.~\eqref{VhatCeta}. Thus, the potential is split between the first and third levels: $\hat{V}$ is added to the 1--1 entry and $V - \hat{V}$ to the 3--3 entry; for $|V| \ll |\Delta_{ee}|$, $\hat{V} \simeq \mcos{13}^{2} V$. At the same time, the 1--2 entry becomes $\tfrac{1}{2}\mcos{\eta}\mtwosin{12}\Delta_{21}$, and a 2--3 entry $\tfrac{1}{2}\msin{\eta}\mtwosin{12}\Delta_{21}$ is generated. Since $\rotmat{13}$ and $\rotmat{\eta}$ are rotations in the same plane, we can define $\theta^{m}_{13} \equiv \theta_{13} + \eta$ and $\rotmat{\theta^{m}_{13}} \equiv \rotmat{13} \rotmat{\eta}$, and write
\begin{equation}
\label{RotHeta}
\rotmat{\theta^{m}_{13}}^{\mathtt{t}} \tilde{H} \rotmat{\theta^{m}_{13}} = \begin{pmatrix}
\lambda_{1} & 0 & 0 \\
0 & \lambda_{2} & 0 \\
0 & 0 & \lambda_{3}
\end{pmatrix} + \frac{1}{2} \mtwosin{12} \Delta_{21} \begin{pmatrix}
0 & \mcos{\eta} & 0 \\
\mcos{\eta} & 0 & \msin{\eta} \\
0 & \msin{\eta} & 0
\end{pmatrix}.
\end{equation}
From Eq.~\eqref{etadef}, $\theta^{m}_{13}$ is the two-flavor mixing angle in matter associated with $\Delta_{ee}$:
\begin{equation}
\label{theta13m}
\cos 2\theta^{m}_{13} = \frac{\varepsilon\left(\mtwocos{13}\Delta_{ee} - V\right)}{\Delta_{h}}, \qquad \sin 2\theta^{m}_{13} = \frac{\varepsilon\,\mtwosin{13}\Delta_{ee}}{\Delta_{h}}.
\end{equation}
 
The two off-diagonal entries in Eq.~\eqref{RotHeta} are of order $\alpha\Delta_{ee}$, but they play different roles. The 2--3 entry is small compared with $|\lambda_{3} - \lambda_{2}|$, which is of order $|\Delta_{ee}|$ for any value of $V$, so it can be set aside even where $\msin{\eta}$ is of order one, close to and beyond the atmospheric resonance. In contrast, $\lambda_{1}$ and $\lambda_{2}$ cross each other at $\hat{V} = \mtwocos{12}\Delta_{21}$, close to the solar resonance, where the 1--2 entry is comparable to $|\lambda_{1} - \lambda_{2}|$. This entry is removed by a second rotation $\rotmat{\theta^{m}_{12}}$ in the 1--2 plane, which leaves $\lambda_{3}$ unchanged. The 1--2 block of Eq.~\eqref{RotHeta} has the form of the two-flavor Hamiltonian in matter for $\Delta_{21}$ and $\theta_{12}$, with $V$ replaced by $\hat{V}$ and the off-diagonal entry multiplied by $\mcos{\eta}$. Hence,
\begin{equation}
\label{theta12m}
\cos 2\theta^{m}_{12} = \frac{\mtwocos{12}\Delta_{21} - \hat{V}}{\Delta_{l}}, \qquad \sin 2\theta^{m}_{12} = \frac{\mcos{\eta}\,\mtwosin{12}\Delta_{21}}{\Delta_{l}},
\end{equation}
with $\Delta_{l}$ given in Eq.~\eqref{ApproxEigenDefs}, where we chose the solution with $\theta^{m}_{12} \to \theta_{12}$ as $V \to 0$. Note that $\theta^{m}_{12}$ is the full 1--2 mixing angle in matter, and not a correction to $\theta_{12}$, because the vacuum 1--2 rotation was not removed in Eq.~\eqref{RotH13}. The new diagonal entries are $\tfrac{1}{2}(\lambda_{1} + \lambda_{2} \mp \Delta_{l})$, which, with $\lambda_{1} + \lambda_{2} = \Delta_{21} + \hat{V}$, are $\zeta_{1}$ and $\zeta_{2}$ in Eq.~\eqref{ApproxEigen}, while $\zeta_{3} = \lambda_{3}$. The 2--3 entry of Eq.~\eqref{RotHeta} is redistributed by this rotation between the 1--3 and 2--3 positions, and we arrive at
\begin{equation}
\label{RotHfinal}
\rotmat{\theta^{m}_{12}}^{\mathtt{t}} \rotmat{\theta^{m}_{13}}^{\mathtt{t}} \tilde{H} \rotmat{\theta^{m}_{13}} \rotmat{\theta^{m}_{12}} = \begin{pmatrix}
\zeta_{1} & 0 & 0 \\
0 & \zeta_{2} & 0 \\
0 & 0 & \zeta_{3}
\end{pmatrix} + \frac{1}{2} \mtwosin{12} \msin{\eta} \Delta_{21} \begin{pmatrix}
0 & 0 & -\sin\theta^{m}_{12} \\
0 & 0 & \cos\theta^{m}_{12} \\
-\sin\theta^{m}_{12} & \cos\theta^{m}_{12} & 0
\end{pmatrix}.
\end{equation}
Since both rotations preserve the trace, $\zeta_{1} + \zeta_{2} + \zeta_{3} = {\Tr}(\tilde{H})$ exactly.
 
The only off-diagonal entries left in Eq.~\eqref{RotHfinal} are proportional to $\mtwosin{12} \msin{\eta} \Delta_{21}$ and couple $\zeta_{3}$ to $\zeta_{1}$ and $\zeta_{2}$, which are separated from it by amounts of order $|\Delta_{ee}|$. Therefore, they modify the eigenvalues only at second order, by terms of order $(\mtwosin{12} \msin{\eta} \Delta_{21})^{2}/|\Delta_{ee}|$, whereas they modify the eigenvectors, and hence the mixing angles in matter, already at first order. Neglecting them, the evolution operator in the intermediate basis becomes
\begin{equation}
\label{ZerothOrderU}
\Utilde(L) \simeq \rotmat{\theta^{m}_{13}} \rotmat{\theta^{m}_{12}} \, \mathrm{diag}(e^{-i\zeta_{1}L}, e^{-i\zeta_{2}L}, e^{-i\zeta_{3}L}) \, \rotmat{\theta^{m}_{12}}^{\mathtt{t}} \rotmat{\theta^{m}_{13}}^{\mathtt{t}},
\end{equation}
and the resulting probabilities are the vacuum ones with $\theta_{13} \to \theta^{m}_{13}$, $\theta_{12} \to \theta^{m}_{12}$, and $\Delta_{kj} \to \DeltaM{kj}$. This zeroth-order approximation coincides with that of Ref.~\cite{Denton2016}, and it is the one used for comparison in Section~\ref{hybridsec}. Unlike the hybrid probabilities, which depend on the neglected entries only through the eigenvalues, the zeroth-order probabilities already depend on them at first order.

\clearpage
\bibliographystyle{JHEP}
\bibliography{Osc_JHEP}
\end{document}